\documentclass{aastex}
\usepackage{emulateapj5}
\usepackage{lscape}
\usepackage{onecolfloat}   %%% If you omit this, then remove the two lines
\newcounter{affil}
\newcommand{\hoggaffil}[2]{%
	\addtocounter{affil}{1}%
	\altaffiltext{\theaffil}{{#2}\label{#1}}}

\newcommand{\hmpc}{\ensuremath{h^{-1}{\rm\,Mpc}}}

\newcommand{\kms}{\ensuremath{\rm\ km\ s^{-1}}}

\newcommand{\spectro}{{\tt idlSpec2d}}

\newcommand{\rpet}{\ensuremath{r_{\rm Petro}}}

\newcommand{\Halpha}{H$\alpha$}
\newcommand{\Hbeta}{H$\beta$}
\newcommand{\Hgamma}{H$\gamma$}
\newcommand{\Hdelta}{H$\delta$}
\newcommand{\NII}{[\ion{N}{2}]}

\newcommand{\OI}{[\ion{O}{1}]}
\newcommand{\OII}{[\ion{O}{2}]}
\newcommand{\OIII}{[\ion{O}{3}]}
\newcommand{\SII}{[\ion{S}{2}]}

\newcommand{\tableskip}{\\[-6pt]}

\begin{document}
\twocolumn[%%% Begin one-column header material

\submitted{Submitted to \textit{The Astrophysical Journal} 8/18/2002}
\lefthead{AVERAGE SPECTRA}
\righthead{EISENSTEIN ET AL}

\title{Average spectra of massive galaxies in the SDSS}
\author{Daniel J.\ Eisenstein\altaffilmark{\ref{Arizona},\ref{Chicago},\ref{HF},\ref{SF}},
        David W.\ Hogg\altaffilmark{\ref{NYU},\ref{IAS}},
	Masataka Fukugita\altaffilmark{\ref{Tokyo}},
	Osamu Nakamura\altaffilmark{\ref{Tokyo}},
        Mariangela Bernardi\altaffilmark{\ref{CMU},\ref{Chicago},\ref{Fermilab}},
        Douglas P.\ Finkbeiner\altaffilmark{\ref{Princeton}},
        David J.\ Schlegel\altaffilmark{\ref{Princeton}},
%%% Builders
	J.\ Brinkmann\altaffilmark{\ref{APO}},
	Andrew J.\ Connolly\altaffilmark{\ref{Pitt}},
	Istv\'an Csabai\altaffilmark{\ref{Eotvos},\ref{JHU}},
	James E.\ Gunn\altaffilmark{\ref{Princeton}},
	\v{Z}eljko Ivezi\'c\altaffilmark{\ref{Princeton}},
	Don Q.\ Lamb\altaffilmark{\ref{Chicago}},
	Jon Loveday\altaffilmark{\ref{Sussex}},
	Jeffrey A.\ Munn\altaffilmark{\ref{Flagstaff}},
	Robert C.\ Nichol\altaffilmark{\ref{CMU}},
	Donald P.\ Schneider\altaffilmark{\ref{PennState}},\\
	Michael A.\ Strauss\altaffilmark{\ref{Princeton}},
	Alex Szalay\altaffilmark{\ref{JHU}},
	Don G.\ York\altaffilmark{\ref{Chicago}}
}

\begin{abstract}
We combine Sloan Digital Sky Survey spectra of 22,000 luminous,
red, bulge-dominated galaxies to get high $S/N$ average spectra
in the rest-frame optical and ultraviolet (2600\,\AA\ to 7000\,\AA).
The average spectra of these massive, quiescent galaxies are 
early-type with weak emission
lines and with absorption lines indicating 
an apparent excess of $\alpha$ elements over solar
abundance ratios.  We make average spectra of subsamples selected by
luminosity, environment and redshift.  The average spectra are
remarkable in their similarity.  What variations do exist in the
average spectra as a function of luminosity and environment are found
to form a nearly one-parameter family in spectrum space.  We present a high
signal-to-noise ratio spectrum of the variation.  We measure the properties
of the variation with a modified version of the Lick index system and
compare to model spectra from stellar population syntheses.  The
variation may be a combination of age and chemical abundance differences, 
but the conservative
conclusion is that the quality of the data considerably exceeds the
current state of the models.
\end{abstract}

]%%% end twocolumn header

\hoggaffil{Arizona}{Steward Observatory, University of Arizona,
		933 N. Cherry Ave., Tucson, AZ 85121}
\hoggaffil{Chicago}{University of Chicago, Astronomy \& Astrophysics Center,
		5640 S. Ellis Ave., Chicago, IL 60637}
\hoggaffil{NYU}{Center for Cosmology and Particle Physics, 
		Department of Physics, New York University,
		4 Washington Place, New York, NY 10003}
\hoggaffil{IAS}{Institute for Advanced Study,
		Princeton, NJ 08540}
\hoggaffil{Tokyo}{Institute for Cosmic Ray Research, 
		University of Tokyo,
		Kashiwa, 2778582, Japan }
\hoggaffil{CMU}{Department of Physics,
		Carnegie Mellon University,
		Pittsburgh, PA 15213}
\hoggaffil{Fermilab}{Fermilab National Accelerator Laboratory,
		P.O. Box 500, Batavia, IL 60510}
\hoggaffil{Princeton}{Princeton University Observatory, Peyton Hall,
		Princeton, NJ 08544}
\hoggaffil{APO}{Apache Point Observatory,
                P.O. Box 59, Sunspot, NM 88349}
\hoggaffil{Pitt}{University of Pittsburgh,
		Pittsburgh, PA}
\hoggaffil{Eotvos}{Department of Physics of Complex Systems, 
		E\"otv\"os University, P\'azm\'any P\'eter s\'et\'any 1, 
		H-1518 Budapest, Hungary}  
\hoggaffil{JHU}{Department of Physics and Astronomy,
                The Johns Hopkins University,
                3701 San Martin Drive, Baltimore, MD 21218}
\hoggaffil{Sussex}{Astronomy Centre, 
		University of Sussex, Falmer, Brighton BN1 9QJ, UK}
\hoggaffil{Flagstaff}{United States Naval Observatory,
		 Flagstaff Station, P.O. Box 1149, Flagstaff, AZ  86002}
\hoggaffil{PennState}{Department of Astronomy and Astrophysics,
                Pennsylvania State University, University Park, PA 16802}
\hoggaffil{HF}{Hubble Fellow}
\hoggaffil{SF}{Alfred P.~Sloan Fellow}

\section{Introduction}

Luminous elliptical and bulge-dominated galaxies are the most massive
galaxies in the Universe.  These objects show little rotation,
have smooth radial profiles, are massive and kinematically hot, have a
narrow range of stellar-mass-to-light ratios, show high metallicities, and
reside preferentially in the Universe's denser environments
\citep[e.g.,][]{Kor89,Rob94,Ben98}.  These galaxies are also the most
intrinsically luminous galaxies in the Universe \citep[e.g.,][]{Tam79,Bla01a}.
That the most optically luminous galaxies are red is remarkable, 
given that unextincted 
red stellar populations generally have higher stellar mass than blue
populations at constant luminosity; it means
that they are considerably more massive than their bluer, fainter spiral
neighbors.

A common paradigm for the formation of bulge-domi\-nated systems
is hierarchical merging of smaller, star-forming progenitors 
\citep{Bar92,Kau96,Bau96}.
In this model, the systems appear old and metal-rich 
because the merging is more common at high redshift and because
the resulting starburst consumes or expels the remaining gas, 
thereby ending star formation \citep{Kau98}.
This dependence upon cosmological merger rates and feedback suggests
that the stellar populations of bulges should display subtle but
important dependencies on mass and environment.

In this paper, we study the aggregate stellar populations of luminous,
red, bulge-dominated galaxies using averaged spectra from the Sloan
Digital Sky Survey 
\citep[SDSS;][]{Yor00}
We select these galaxies because we know
that they dominate the stellar mass density of the Universe
\citep{Fuk98,Hog02} and because they show great regularities in their
properties \citep[e.g.,][]{Fab73,Vis77,Djo87,Dre87,Kor89,Bow92,Rob94, Ber01}.  
In addition, the great age
of their stellar populations ease some, but not all, of the difficulties
in understanding superimposed younger populations
\citep{Fab72,OCo76,Gun81,Cha96}.

As the luminosity-weighted average spectrum does measure the 
total stellar population of an entire sample of galaxies, it
is a well-defined mean property that can be compared to theoretical
models of galaxy formation \citep[][e.g.,]{Bal02}.
We focus not on the interpretations of the average spectrum itself,
but rather on the variations of this average with luminosity,
environment, and redshift.  These variations should be able to test
predictions that more luminous (or massive) galaxies have higher
metallicity and galaxies in clusters have older stellar populations.
Previous observational studies \citep[e.g.,][]{Bow90,Guz91,Ros94,Jam99,Ter99,Pog01}
comparing spectra of individual galaxies have found evidence to support 
these hypotheses.  
Of course, one would
also expect to find that galaxies at high redshift should have younger
stellar populations
than similar galaxies at low redshift, but even this trend can be altered if
massive, bulge-dominated galaxies are still forming at low
redshift \citep{Fra95, van01}.

Because we are interested in comparing the average spectra of samples, 
it is critical that the samples be selected so that the 
variation between the samples is as controlled as possible.
With SDSS data, we can select uniform samples according to rest-frame
photometry and morphological properties across a wide range of environments
and redshifts,
allowing trends in the data to be attributed to astrophysics rather
than selection biases.

The average spectra are of extremely high signal-to-noise ratio, thereby
allowing analyses that would be impractical on individual SDSS spectra.
Over 100 absorption lines are apparent in the composite.  
Indeed, the resulting spectra exceed the ability of the best spectral 
models of stellar populations to interpret, 
especially given the non-solar ratio of $\alpha$-element to iron-peak abundances
\citep{Wor92,Gon93,Dav93,Paq94,Wor98,Pro02} apparent in the data.

The outline of this paper is as follows.  In \S~2, we describe the 
relevant aspects of SDSS data and introduce the methods for our analysis,
including principal component analysis (PCA) and a modification of the 
Lick absorption line index system.  In \S~3, we describe our sample 
selection and environment estimation.    We present a general analysis 
of the average spectra in \S~4 and analysis of the variations with
luminosity and environment in \S~5.  Variations with redshift are
presented in \S~6.  In \S~7, we use the highest redshift sample to 
determine the average spectrum in the mid-ultraviolet.
We conclude in \S~8.

Except where noted otherwise, we adopt a conventional cosmological
world model with $H_0=100\,h~{\rm km\,s^{-1}\,Mpc^{-1}}$ and
$(\Omega_M,\Omega_{\Lambda})=(1/3,2/3)$.

\section{Data and Procedures}

\subsection{SDSS Imaging}
\label{sec:imaging}

The SDSS is obtaining $u$, $g$, $r$, $i$ and $z$-band drift-scan images
of the Northern Galactic Cap and spectra of roughly $10^6$
galaxies in that region \citep{Yor00}.  The photometric system
and imaging hardware are described in detail elsewhere \citep{Fuk98,
Gun98, Hog01, Smi02}.  An automated image-processing system detects
astronomical sources and measures photometric and astrometric
properties \citep{Lup01a, Sto02, Pie02, Lup02} of them.  
Finally, galaxies are selected for spectroscopy by two
algorithms \citep{Str02, Eis01} described further in \S~\ref{sec:select}.

The photometric parameters of interest for this study are as follows.
The \citet{Pet76} radius $\theta_{r,\mathrm{pet}}$ is the angular radius at
which the mean surface brightness of the source in the SDSS $r$-band
image inside that radius is five times higher than the mean surface
brightness in a narrow annulus centered on that radius.  The Petrosian
flux (with corresponding magnitude $m_\mathrm{pet}$) is the total flux
(in any of the five SDSS bandpass images) within a circular aperture
of radius $2\,\theta_{r,\mathrm{pet}}$ (twice the Petrosian radius in
the $r$-band).  The half-light and 90\% radii $\theta_{50}$ and
$\theta_{90}$ are the angular radii within which 50\% and 90\% of the
$r$-band Petrosian flux is found.  The concentration $c$ is defined to be the
ratio $\theta_{90}/\theta_{50}$.  The de~Vaucouleurs likelihood
$L_\mathrm{dev}$ is the likelihood that the galaxy's $r$-band 2-dimensional image
was generated by a true de~Vaucouleurs
elliptical profile convolved with the seeing.  The exponential likelihood
$L_\mathrm{exp}$ is a similar likelihood of an exponential profile.
The model magnitude $m_\mathrm{model}$ is the total magnitude
corresponding to the best-fit profile, which is the best-fit de~Vaucouleurs
profile for all galaxies in this study.  All
magnitudes are corrected for extinction using the 
\citet{Sch98} predictions.

Two samples of galaxies are selected from the SDSS imaging for spectroscopy.
The larger sample, with about 88\% of the allotment, 
is a flux-limited sample of galaxies that extends
to $\rpet<17.77$ \citep{Str02}.  
We refer to this as the MAIN sample.
The smaller portion, with the remaining 12\% of the allotment, 
uses two color-magnitude cuts to seek
luminous, early-type galaxies that are fainter than the MAIN
limit.  This sample is known as the luminous, red galaxy (LRG)
sample.  The LRG sample is roughly volume-limited to $z\approx0.4$
and contains additional galaxies to $z\approx0.55$ with a flux limit
of $\rpet=19.5$.  Further details are found in \citet{Eis01}.

\subsection{SDSS Spectroscopy}

The SDSS uses two fiber-fed double-spectrographs to measure spectra
of objects from 3800 \AA\ to 9200 \AA\ with a resolution of about 1800.
Each plug plate holds
640 fibers, yielding 608 spectra of galaxies,
quasars, and stars and 32 sky spectra per pointing \citep{Bla01b}.  
Fibers cover $3''$ diameter circular apertures on the sky.  
Each spectrograph handles 
320 fibers (which we call a half-plate), and the two halves are 
reduced independently.  

In detail, the instrumental resolution and pixel scale are close to 
constant in logarithmic wavelength rather than wavelength.  The
instrumental resolution is about $170\;\kms$ FWHM.  The \spectro\ 
software pipeline (Schlegel, Burles, et al., in prep)
combines multiple exposures of a given object
and resamples the total spectrum onto a grid of wavelengths that
is logarithmically-spaced by $69.1\;\kms$.

The flux calibration procedure is summarized in \citet[\S~3.3 \& 4.10.1]{Sto02} and
will be fully detailed in Schlegel, Burles, et al. (in prep).
The flux calibration is imposed on each plate by a set of 8 spectrophotometric
standard stars, chosen by color to be F subdwarf stars.
The SDSS does not use an atmospheric refraction corrector, so 
the effective fiber position on the sky shifts slightly as a 
function of wavelength.  In the presence of brightness gradients,
this creates a fluxing error.  This is corrected by refering the 
broadband spectrophotometry to $5''\times8''$ aperture ``smear'' 
exposures.  As early-type galaxies have small color gradients, we
expect that this smear correction will remove the effects of 
atmospheric refraction.

The spectrophotometric calibration appears to be correct in the average to
better than 10\%.  Since each average spectrum involves galaxies
from a moderate range of redshifts, systematic wavelength-dependent
spectrophotometry biases in the average spectra will be strongly 
suppressed on scales below a few hundred \AA.
Comparing our average spectra in different redshift
bins (see \S\ \ref{sec:lrg}) 
shows differences of under 10\% (4\% {\it rms}), some of which may
be due to actual evolution.  As this involves comparison of spectrophotometry
at different observed wavelengths, we infer that the mean spectrophotometry
is accurate at this level, up to perhaps an overall tilt.
We will return to this point in \S\ \ref{sec:onedim}.
Recent efforts (Tremonti \& Schlegel, private communication)
suggest that the current calibrations are slightly too blue
(roughly 10\% in flux over the spectral range)
with additional residuals blueward of 4300\AA (observed frame).

Redshifts are found by comparing to stellar templates
(Frieman et al., in prep.; Schlegel et al., in prep.).
Repeat observations show that the errors in the redshift estimates 
are less than $30\;\kms$ for MAIN sample galaxies and degrade to about
$100\;\kms$ for the highest redshift LRGs.

\subsection{Spectroscopic Sample}

We use spectra from 261 separate plates; a total sample of about
120,000 galaxies.  Certain plates have been 
observed more than once, and a small fraction of objects appear on more than
one plate (for quality assurance).  We use only one spectrum of
each object, chosen by the best median signal-to-noise ratio on each
candidate half-plate.  Only plates that meet survey quality standards
are used.  

A few spectra were found to contain sufficiently serious reduction 
failures as to dominate the principal component analysis; we rejected
the entire plate in this case (so as not to skew the average by
some undetected pattern among the corrupted objects).
This eliminated 12 plates, although some of these had other 
exposures that could have been used. 

We select samples of red, bulge-dominated galaxies from these plates; the
selection is described in \S~\ref{sec:select}.

\subsection{Principal component analysis}

We are interested in average spectra; however, we choose to 
go further and perform principal component analyses (PCA) on each
set of spectra.
This isolates a set of eigenspectra ranked by the amount
of the total variance in the input set that each explains.
PCA is useful for identifying variations in the spectra beyond the mean,
and we find that emission line components are isolated by this technique.  

We can recover the average spectra from the PCA decomposition
by finding the mean of each PCA coefficient and summing the eigenspectra
weighted by these means.  
Were we to include all the eigenspectra, the resulting linear combination
would be exactly the average of the spectra used as input to the PCA.
However, we include only the first 20 eigenspectra.
We have verified that this procedure makes no difference relative
to the true average; the means of the PCA coefficients converge
rapidly to zero.  One can see that the truncation does
not constitute a smoothing of the spectra by noting that permuting
the order of the pixels within the definition of the spectral space
would not affect the PCA results.  

We begin with spectra in $F_\lambda$ units, but we 
use the redshifts
to rescale each source to its intrinsic luminosity spectrum $L_\lambda$.
This does not fully reproduce the mean spectrum of a 
volume-limited set of galaxies, but in practice the luminosity
range of our samples are sufficiently narrow---0.5 mag---that 
the corrections to a volume-limited mean are negligible.
To close approximation, each average spectrum will represent 
the volume-averaged stellar population of galaxies with the
chosen intrinsic properties.

Before performing the PCA step, we pre-condition the spectra.
We shift each spectrum to the rest frame according to its redshift.
We do not include shifts smaller than a unit spacing of
our wavelength grid ($69.1\;\kms$).  This is equivalent to adding a 
small additional velocity dispersion to the sample; however, the
effect is only $20\;\kms$ ($69\div\sqrt{12}$, the standard deviation of a boxcar), 
which is small compared to the actual velocity
dispersions of the galaxies.  We then select a particular range
of rest-frame wavelengths, say 3650\AA\ to 7000\AA,
to ensure that the full range of wavelengths is covered by SDSS
spectra for the full range of redshifts included in a particular sample.

We iteratively interpolate over pixels for which \spectro\ has set 
warning mask flags.  We begin by linear interpolating over all masked
pixels and computing the mean spectrum.  We then return to the masked pixels,
interpolate with the mean spectrum times a linear function,
and recompute the mean.  We repeat
this step four times to ensure convergence.
We increase the masked region around the 
sky features at 5577\AA, 5890\AA, 5896\AA, 6300\AA, 6364\AA, 7245\AA, and 7283\AA,
typically to 10\AA, so as to avoid rare problems of poor sky subtraction.
Since the samples involve a wide range of redshift, masking at
a fixed observed wavelength enters at different rest-frame wavelengths
and therefore has little effect on the average spectrum.
We remove entirely any objects that have more than 20 pixels 
masked for missing data (not counting pixels in the first and last
50 pixels of a given spectrum).

We subtract the continuum of each individual spectrum by removing 
the first $N_F$ sinusoidal components.  We pick $N_F=8$ for our 
3650\AA\ to 7000\AA\ stacking, $N_F=6$ for 3500\AA\ to 6000\AA\ 
stacking, and $N_F=5$ for 2600\AA\ to 4400\AA\ stacking.  
We have tried other choices with little change.  For example,
we have tried extracting narrow wavelength regions with no 
continuum subtraction and get the same answers.

Because we subtract the continuum from our spectra before performing
the PCA, our eigenspectra lack a continuum.  We restore this continuum 
after the PCA step by adding back the average
of the subtracted continua.  This procedure exactly reproduces the mean spectrum
but protects the PCA from being dominated by broad-band variations.
When comparing spectra from different samples, we rescale to a common
overall amplitude but do not otherwise adjust the continua.

Before performing the PCA, we renormalize the different wavelengths by 
the inverse of the average error
in that pixel.  This doesn't affect the average spectrum but does affect
subsequent components.  In practice, the weighting is mild.
After performing the PCA, or more exactly, the diagonalization of the
product of the matrix of spectra with its transpose, we restore the 
original physical normalization of the wavelengths in each eigenspectrum.

In practice, we find that a single component dominates the spectrum of
eigenvalues.  This eigenspectrum shows a familiar early-type
absorption line spectrum.  Because the ratio of this eigenvalue
to the others is so large, the first component can deviate only slightly
from the mean of the dataset.

Regarding different samples, it is important to stress that the
entire PCA process has been applied to the two sets of galaxies completely
independently.  In particular, the comparison of average spectra is
not constrained to differ only in a low (20) dimensional subspace
of retained eigenspectra.

\subsection{Error estimation}

To construct an estimate of the error on the average spectrum,
we divide each sample into two, find the average spectrum
of each half, and analyze the rms differences between the two halves.
Before differencing, we remove 20 low-pass sinusoidal modes (from 
over 2000 pixels) to be sure that the continuum is
not affecting our estimate of the small-scale noise.  
There is essentially no structure apparent in the differenced spectra.
Because there are small correlations between neighboring pixels
(e.g.\ because of the sub-pixel wavelength shifts between exposures),
we smooth the difference with a box car filter of 5 pixel width
and multiply the result by $\sqrt{5}$.  Were the pixels independent,
this would not matter, but for correlated pixels this smoothing creates the
residual that should apply to multi-pixel averages such as are
found in spectral indices.
We form the square of the smoothed difference spectrum
and then smooth with a 100 pixel box-car.  This effectively
creates the variance of the differenced spectrum as a function of wavelength.

The galaxies are split into the two halves according to whether their 
plate number is even or odd.  This means that reduction errors that
might affect an entire plate are included in the noise estimate,
because all of the affected galaxies will go into one half or the other.
Conversely, any secular changes in target selection algorithm or imaging properties
are isolated from the error analysis.  If we had divided the sample
into early plates and the late plates, small differences in target selection
(perhaps simply caused by different photometric calibrations in
different parts of the sky)
could have resulted in artificial differences between the mean spectra.

If one has two spectra of equal signal-to-noise ratio, then the
noise in the difference is twice the noise in the average.
Hence, as our even and odd spectra
should have quite similar signal-to-noise ratio, we estimate the variance
in the average spectra as being one quarter of that in the 
smoothed square difference.

Although the noise in a single spectrum 
can be a complicated and spiky function of wavelength due to 
the presence of sky lines and detector flaws, the noise in the combined
spectrum should be a smooth function of wavelength.  Noise features
at a narrow observed wavelength range are smeared out by the superposition of many 
different redshifts, while detector issues such as bad CCD columns are
eliminated by the superposition of spectra from many different positions in
the detector.  Only the broad variations in response across wavelength
(e.g., the drop in sensitivity shortward of 4000\AA) should survive 
to affect the noise in the stacked spectrum.

The error at certain rest-frame wavelengths is dominated not by photon noise but
rather by the shot noise of the presence of rare galaxies in the sample.
In particular, a small fraction of our selected galaxies have emission lines,
and the presence or absence of a few strong emission-line galaxies can affect
the results at the line wavelengths.  We could treat this by bootstrap
resampling of the input samples, but the line emissions are sufficiently
weak that we will focus only on the isolation and subtraction of the lines 
and not on the error in that subtraction.

The signal-to-noise ratios of the average spectra are very high, 
typically a few hundred per pixel (see Table \ref{tab:summary}).
Because our error estimation is based on internal differences, it
does not account for errors that are common to the two half samples.
Generally, systematic effects are functions of observed wavelength,
which means that their impact will
be smeared out by the combination of objects of different redshifts.
This suppresses such error modes on scales less than a few hundred
\AA.  We have few ways at present to assess the errors on larger scales, but
the differencing of samples at very different redshifts do suggest
that the errors are small (4\% {\it rms}).
We are not aware of any systematic effects that would be functions of
rest-frame wavelength.  We will present another validation of the
quoted errors in \S~\ref{sec:onedim}.

\subsection{Comparison between spectra and to models}
\label{sec:vazdekis}

The galaxies in the samples to be defined in \S\ \ref{sec:select}
differ in their velocity dispersions.
In order to make fair comparisons among the average
spectra, we therefore smooth all the average spectra to a common velocity
dispersion of $\sigma_v= 325~\mathrm{km\,s^{-1}}$, which
is the largest velocity dispersion of the individual samples.  

In order to interpret our average spectra in terms of age, 
metallicity, or other variations, we need to compare our spectra
to model spectra of stellar population synthesis models.
We use the models of \citet{Vaz99} (hereafter V99), who
provides 2\AA\ resolution spectra of single-age, single-metallicity,
solar-element ratio stellar populations.  Because the 
resolution of these models is better than the resolution 
achieved in galaxies with large velocity dispersion, we can
treat these model spectra identically to our data and thereby
do fair comparisons between the two.  We use single power-law
initial mass function V99 models throughout.

\subsection{sLick indices}
\label{sec:sLick}

We measure absorption line strengths by a revision of the most
recent version of the ``Lick'' system \citep{Fab85, Wor94, Tra98}.
In addition to the standard Lick indices, we have included 4 indices
for measuring the \Hgamma\ and \Hdelta\ absorption lines \citep{Wor97}.
The Lick system involves measuring line fluxes in narrow wavelength
ranges relative to a ``pseudocontinuum'' found by linearly
interpolating between two spectral averages on either side of the
lines.  

We cannot match the fluxing and calibration of the Lick system 
but have attempted an intermediate step of matching the 
wavelength-dependent resolution of the Lick IDS instrument
\citep[using a linear interpolation of the table in][]{Wor97}.  
The highest velocity dispersions of
any of our spectra is $325 \kms$ at SDSS resolution.  
This corresponds to $275 \kms$ at 6000\AA\ as smoothed to 
Lick IDS resolution.  Hence, when computing indices, we smooth our 
spectra to match, at each wavelength, an intrinsic dispersion of $275\kms$ 
at the Lick IDS resolution.
For example, this corresponds to $450\kms$ at 4000\AA.
The result is the spectrum of a galaxy with $275\kms$ dispersion 
as it would be observed by the Lick IDS instrument.
Because, strictly speaking, the Lick system is defined
at zero velocity dispersion (at Lick IDS spectral resolution), we
refer to the
indices used in this work, normalized to $275~\mathrm{km\,s^{-1}}$, as
smoothed-Lick or ``sLick'' indices.

We introduce the sLick indices rather than work on the true Lick system
because (a)~The Lick IDS spectra were referenced to a tungsten
blackbody for overall calibration; (b)~the indices are, strictly
speaking, calibrated to a defined set of standard stars, which have
not been observed with the SDSS telescope; and (c)~the Lick system is
defined at zero velocity dispersion, at which no galaxies can ever be
observed, although some empirical corrections for velocity dispersion
have been defined \citep{Tra98}.  For all these reasons, direct
comparison of our sLick index measurements (in
Table~\ref{tab:slickellip1}) to measurements in the Lick system found
in the literature must be made with care.

On the other hand, we have treated the V99 model spectra in an identical
way to the observed spectra,
so the internal consistency of comparing our observed indices with
the models is reliable.
For the same reason, our measurements of differences or
derivatives of sLick indices among our average spectra given in 
Table~\ref{tab:slickellip1} are accurate.

\section{Massive galaxy samples in the SDSS}
\label{sec:select}

\subsection{Morphology and color selection in the MAIN sample}
\label{sec:main}

\begin{figure}[tb]
\noindent\centerline{\epsfxsize=\columnwidth\epsffile{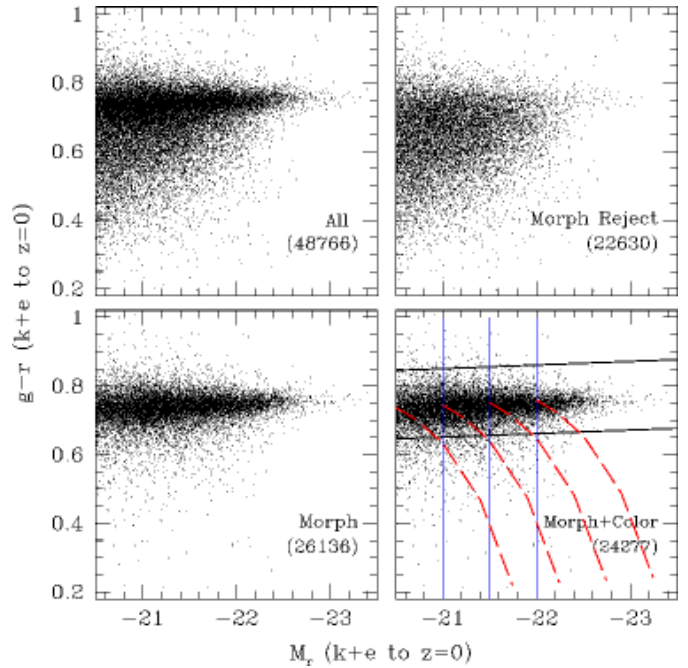}}
%%% \plotone{ellip_cuts.eps}
\caption{\label{fig:mainsel}%
Rest-frame $g-r$ color versus absolute $r$-band magnitude 
(both corrected for passive
evolution) of MAIN galaxies.  The effects of the morphological radial profile
cuts is shown
in the bottom two panels.  Galaxies failing the morphology cut are in
the upper right panel.
The lower right panel shows the color cuts and luminosity bins.
The dashed lines show the locus of color and magnitude of an old
population with a fading 20\% young single-age burst \citep{Bru01}.
We stress that the color cut is excluding many
fewer galaxies at fixed stellar mass than at fixed luminosity.
}
\end{figure}

\begin{figure}[t]
\noindent\centerline{\epsfxsize=\columnwidth\epsffile{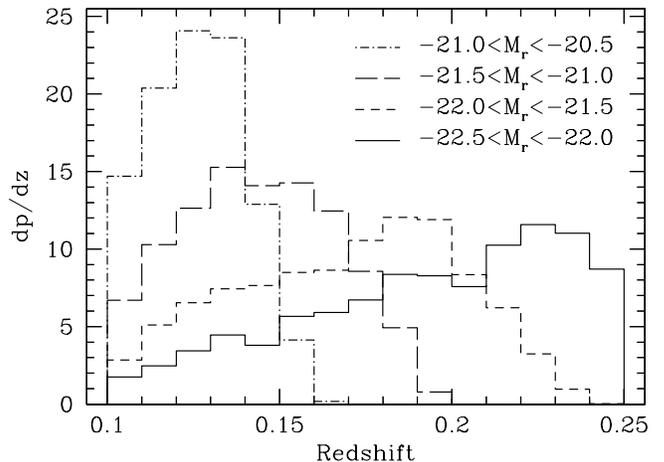}}
\caption{\label{fig:redshift}%
The redshift histograms for the MAIN galaxy luminosity-binned samples.
}
\end{figure}

The MAIN sample contains galaxies of all types, but we would like
to select massive, red, bulge-dominated galaxies.  
We do this selection primarily on the basis
of morphology but cannot avoid the need to remove color outliers.

We consider 4 luminosity bins within the MAIN sample of galaxies:
$-20.5<M_r<-21$, $-21<M_r<-21.5$, $-21.5<M_r<-22$, and $-22<M_r<-22.5$.
These absolute magnitudes are $k+e$ corrected,
de-reddened, $r$-band Petrosian
magnitudes.  We use the bandpass and evolution ($k+e$)
correction of a passively evolving old stellar population \citep{Eis01} 
to adjust the galaxy photometry to $z=0$.
Since the redshift range of each bin is small and since the evolution 
correction is only 1 mag per unit redshift, the assumed evolution 
correction is not important for the results but merely serves to 
compare the galaxies to those at $z=0$.
Note that $M^*_r \approx -20.8$ from the SDSS luminosity function \citep{Bla01a},
so the least luminous bin has galaxies whose luminosity is roughly $L^*$,
while the most luminous bin has galaxies four times more luminous.

We make two purely morphological cuts on the sample:
1) we require that the inverse concentration $1/c$ be less than 0.37, and
2) we require that the de~Vaucouleurs model
fit be 20 times more likely than the exponential fit 
(excluding galaxies for which both likelihoods underflow to zero).  
See \S~\ref{sec:imaging} for the definitions of these statistics.
Both of these cuts favor the traditional 
surface brightness profiles of elliptical galaxies and bulges.  
We do not perform cuts on the asymmetries or residual lumpiness of 
the galaxy.  

We find that these radial profile cuts do significantly alter the distribution
of galaxies in a rest-frame color versus absolute magnitude plot.
The red sequence of early-type galaxies is strongly favored.
Based on similar samples in \citet{Shi01}, \citet{Stra01}, and \citet{Hog02}, 
we expect that more than 2/3 of the sample are of E/S0 morphology, 
with the remainder being early-type spirals.

Figure \ref{fig:mainsel} shows the distribution of galaxies in rest-frame
color versus absolute magnitude with and without the morphological cuts.
A small fraction of the radial-profile-selected galaxies 
lie well below the red sequence in $g-r$.  If one considers slices at
constant $r$-band luminosity, this fraction of blue galaxies is $\sim\!5\%$.
It should be stressed that the fraction of blue galaxies in slices of
constant stellar mass would be considerably lower.
Blue galaxies fade as they age and redden.  Hence, one should
compare the number of lower luminosity galaxies on the red sequence
to the number of higher luminosity blue galaxies.  
In Figure \ref{fig:mainsel}, 
we display the loci of magnitude and color populated
by a burst of 20\% of new stars on top of an old population,
using models from \citet{Bru01}.  Following these curves
rather than lines of constant luminosity, we would say that very few massive
galaxies have been excluded by our color cut.  In this model,
20\% bursts require about 2 Gyr to redden enough to pass our color cut.

As we are interested in massive quiescent galaxies, we decide to cut these
blue outliers using a color cut.  We fit the rest-frame color-magnitude
relation to a line of $(g-r)_0 = 0.74-0.014 (M_r+20)$ and keep only 
those galaxies that fall within 0.1 mag in rest-frame color (region shown
in Figure \ref{fig:mainsel}).
This excludes 3\% of the galaxies in the highest luminosity bin
and 8\% in the lowest (see Table \ref{tab:summary}).

Our color selection obviously excludes vigorously 
star-forming galaxies from the sample.  Hence, we cannot claim that the 
average spectrum we find represents the average light from {\it all} 
bulge-dominated systems.  The average age of the sample
will be biased towards older ages (although of course the dominant
problem in interpretation of optical spectra is the opposite, 
namely that a few young stars will bias the inferred age low).

We would prefer to compare galaxies at equal redshifts, and therefore
we restrict our sample to redshifts between 0.1 and 0.25.
However, our lower luminosity subsamples are drawn from a 
characteristically lower portion of this range, so there may
be a small amount of passive evolution mixed into the spectral comparison
across luminosity.  Histograms of the redshift distribution are
shown in Figure \ref{fig:redshift}.

Table \ref{tab:summary} lists the sample size, mean redshift, mean luminosity,
selection properties, and average spectrum signal-to-noise ratio 
for the four luminosity samples (see the rows with environment ``All''), 
as well as for the environment and redshift samples discussed in the 
next two sections.

\begin{table*}[tb]\footnotesize
\caption{\label{tab:summary}}
\begin{center}
{\sc Summary of Sample Properties\\}
\begin{tabular}{ccccccccc}
\tableskip\tableline\tableline\tableskip
Magnitude & Environment & Number & Used & S/N & $\left<z\right>$ & $\left<M_r\right>$ & $\sigma$ applied & Color reject \\
\tableskip\tableline\tableline\tableskip
$-22.5<M_r<-22.0$ & All        & 2568 & 2394 & 487 & 0.193 & -22.19 & 166 &  72 \\
$-22.5<M_r<-22.0$ &  $0\le N\le 3$ &  356 &  326 & 159 & 0.197 & -22.17 & 170 &  21 \\
$-22.5<M_r<-22.0$ &  $4\le N\le 9$ &  822 &  778 & 323 & 0.195 & -22.19 & 156 &  11 \\
$-22.5<M_r<-22.0$ &    $N\ge 10$ &  531 &  494 & 260 & 0.186 & -22.22 & 134 &   6 \\
\tableskip\tableline\tableskip
$-22.0<M_r<-21.5$ & All        & 5830 & 5412 & 464 & 0.170 & -21.73 & 181 & 276 \\
$-22.0<M_r<-21.5$ &  $0\le N\le 3$ & 1253 & 1148 & 303 & 0.171 & -21.72 & 198 &  77 \\
$-22.0<M_r<-21.5$ &  $4\le N\le 9$ & 1905 & 1773 & 400 & 0.170 & -21.73 & 192 &  77 \\
$-22.0<M_r<-21.5$ &    $N\ge 10$ &  732 &  686 & 256 & 0.166 & -21.75 & 166 &  18 \\
\tableskip\tableline\tableskip
$-21.5<M_r<-21.0$ & All        & 6991 & 6477 & 697 & 0.144 & -21.25 & 200 & 414 \\
$-21.5<M_r<-21.0$ &  $0\le N\le 3$ & 1784 & 1633 & 385 & 0.143 & -21.25 & 223 & 116 \\
$-21.5<M_r<-21.0$ &  $4\le N\le 9$ & 2166 & 2009 & 414 & 0.144 & -21.25 & 210 & 128 \\
$-21.5<M_r<-21.0$ &    $N\ge 10$ &  648 &  603 & 264 & 0.142 & -21.25 & 199 &  22 \\
\tableskip\tableline\tableskip
$-21.0<M_r<-20.5$ & All        & 5111 & 4749 & 626 & 0.126 & -20.78 & 231 & 424 \\
$-21.0<M_r<-20.5$ &  $0\le N\le 3$ & 1425 & 1319 & 350 & 0.125 & -20.77 & 237 & 130 \\
$-21.0<M_r<-20.5$ &  $4\le N\le 9$ & 1446 & 1343 & 395 & 0.126 & -20.78 & 223 & 112 \\
$-21.0<M_r<-20.5$ &    $N\ge 10$ &  488 &  463 & 205 & 0.125 & -20.77 & 228 &  25 \\
\tableskip\tableline\tableline\tableskip
Magnitude & Redshift & Number & Used & S/N & $\left<z\right>$ & $\left<M_g\right>$ & $\sigma$ applied & \\
\tableskip\tableline\tableline\tableskip
$M_g<-21.8$ & $0.15<z<0.25$ &  297 &  276 & 184 & 0.210 & -21.98 &  90 & \\
$M_g<-21.8$ & $0.25<z<0.35$ &  685 &  641 & 184 & 0.305 & -21.99 &  51 & \\
$M_g<-21.8$ & $0.35<z<0.45$ & 1286 & 1200 & 169 & 0.403 & -22.00 &  71 & \\
$M_g<-21.8$ & $0.45<z<0.50$ &  762 &  700 &  87 & 0.475 & -22.01 & 139 & \\
\tableskip\tableline\tableline\tableskip
\end{tabular}

\end{center}
NOTES.---%
Column titled ``Number'' lists the number of objects that pass our morphology,
luminosity, density, and color cuts.  
The column titled ``Color reject'' lists the number
that passed the morphology, luminosity, and density cuts, but failed the
color cut.
The column titled ``Used'' lists the number
actually used; some objects fail our spectral quality cuts on the amount of 
missing or masked data.  
The column titled ``S/N'' lists the average signal-to-noise ratio per pixel
of spectrum (roughly 1 \AA) in the stacked spectrum.
The average redshift and absolute magnitude in each sample is given,
as is the dispersion of the Gaussian required to smooth the average spectrum
to a velocity dispersion of $325\;\kms$.
The sums of the 3 environment samples at a given luminosity do not equal
the ``All'' environment sample because we exclude galaxies from the 
environment sample that fall less than $0.5\hmpc$ from the edge of the 
survey imaging.
\end{table*}

\subsection{Luminous LRG selection}
\label{sec:lrg}

\begin{figure}[tb]
\noindent\centerline{\epsfxsize=\columnwidth\epsffile{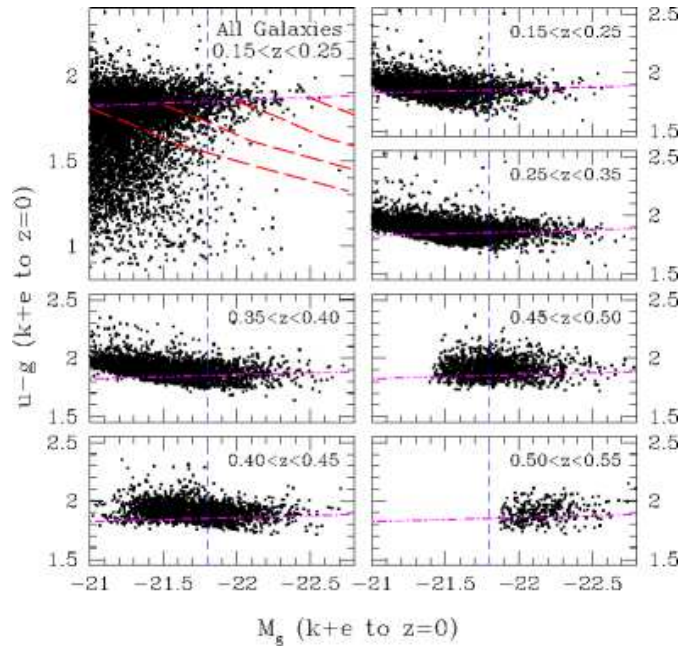}}
%%% \plotone{lrg_cuts.eps}
\caption{\label{fig:lrgsel}%
The rest-frame color versus magnitude (both corrected for passive evolution)
distribtion of 
({\it upper left panel}) all galaxies at $0.15<z<0.25$ and
({\it other panels}) LRGs in different redshift slices.
The horizontal dot-dashed line is a fit to the color-magnitude
locus of the red sequence.  The full LRG sample would result
in a bias towards galaxies on the red side of the red sequence.  
By imposing an additional magnitude cut (vertical dashed line), 
we can include the whole red sequence.  
The sample is volume-limited to $z\approx0.5$.
The diagonal dashed lines are fading bursts as in Figure \protect\ref{fig:mainsel}.
}
\end{figure}

\citet{Eis01} show that the LRG spectroscopic sample contains luminous
galaxies with red rest-frame colors and early-type spectra out to $z\approx0.55$.
The sample is roughly volume-limited at $z<0.4$; however, the selection
does create an oblique dependence on luminosity
and rest-frame color.  Intrinsically bluer galaxies must be more luminous
than their redder cousins to enter the sample.

Luminous galaxies drawn from the MAIN sample at $0.15<z<0.2$, where there
is no color selection, show an obvious ridge in color-magnitude space
from the prevalence of early-type galaxies.  At the highest luminosities,
the red sequence is particularly obvious because the frequency of galaxies
drops significantly blueward of the sequence.  We would
like to select all of the galaxies on the red sequence.

This red sequence is quite
narrow, but not infinitely so.  Hence, we have the problem that for some
range of luminosities, galaxies on the red side of the red sequence would
qualify for the LRG sample while those on the blue side would not. 
To gather a sample for which the entire red sequence will be included,
we must use an even higher luminosity cut than the normal LRG selection criteria.

We therefore require $M_g<-21.8$ within the LRG sample.  This luminosity
refers to the passively evolved, rest-frame $g$-band luminosity 
$k$-corrected from the $r$-band Petrosian magnitude, 
as described in \citet{Eis01}.  For comparison to our samples from 
MAIN, we note that this cut corresponds to $M_r\lesssim-22.55$.
As shown in Figure \ref{fig:lrgsel}, this cut allows the LRG
sample to include the full red sequence.
In addition, the higher luminosity cut means that the 
sample, which would otherwise be subject to flux limitations at $z\gtrsim0.40$,
retains its volume-limited nature until $z\approx0.5$.  

We limit the LRG selection to $0.15 < z < 0.50$ and divide this into
three redshift bins 0.1 in width plus a fourth bin at $0.45<z<0.50$.  
The limit at low redshift is primarily
due to the small number of sufficiently luminous galaxies in the small
closer volume.  Similarly, there are not many galaxies in the LRG sample
at $z>0.50$, and these would not be volume-limited to $M_g=-21.8$.

\subsection{Local density estimation}
\label{sec:density}

We wish to study the spectral properties of our galaxies as a function of
their environment.  As our spectra are of luminous galaxies, we do not
have full spectroscopic coverage of the fainter neighbors.  We therefore
resort to counting nearby galaxies from the SDSS imaging data.  This
has the advantage that it is independent of the SDSS spectroscopic 
coverage, which of course is only partially complete, and that the
same luminosity cuts can be applied to all galaxies.
To include a poor-man's photometric redshift scheme, we restrict the count
to galaxies consistent with early-type colors at the appropriate redshift.

Given the spectroscopic redshift of a particular luminous galaxy, 
we calculate the angular scale
corresponding to a transverse distance of $0.5\hmpc$, the $r$ band
luminosity of an $L^*$ passively-evolved early-type galaxy, and 
the $g-i$ color of that galaxy.  The last of these is based on the 
average color of the two evolving models in \citet{Eis01}.
We then count the number of galaxies that fall within 0.15 mag in $g-i$
model color, within 1 mag brightward and 2 mag faintward in $\rpet$ flux,
and between $0.03$ and $0.5\hmpc$ in transverse distance.  
Primary galaxies that fall within $0.5\hmpc$ of the edge of the imaging sample 
are excluded from the environmental studies.  This means that the
samples for environment variations are smaller than those for luminosity
variations (see Table \ref{tab:summary}).

This count of nearby early-type galaxies gives us an estimate of the 
local environment of our candidate.  However, the statistic is noisy.
With random positions inside the survey region, 
we find an average of 3.6 neighbors,
nearly independent of redshift.
The average number of neighbors around the actual galaxies increases
with luminosity: 5.8, 5.7, 6.4, and 8.95 for luminosity bins of 
$-21.0 < M_r < -20.5$, $-21.5 < M_r < -21.0$, $-22.0 < M_r < -21.5$,
and $-22.5<M_r < -22.0$, respectively.  Hence, while we have clear evidence
that galaxies are clustered, the Poisson error on the background
is not small relative to the average signal.  We cannot make fine
distinctions about the local density but instead can construct
generously large bins that should be monotonic in their typical 
density.

For each luminosity bin in the MAIN sample, we divide the sample into 
three density bins---those 
with 10 or more neighbors,
those with 4 to 9 neighbors, and those with 3 or fewer neighbors---for
a total of 12 samples.
While the dividing line between the samples will be blurred, with
equivalent galaxies jumping the boundary due to background Poisson
fluctuations, bins this large should be reasonably distinct as
regards the actual (non-projected) density of the included galaxies.

\section{The Average Spectra}

\begin{figure*}[tb]
\centerline{\epsfysize=4in \epsffile{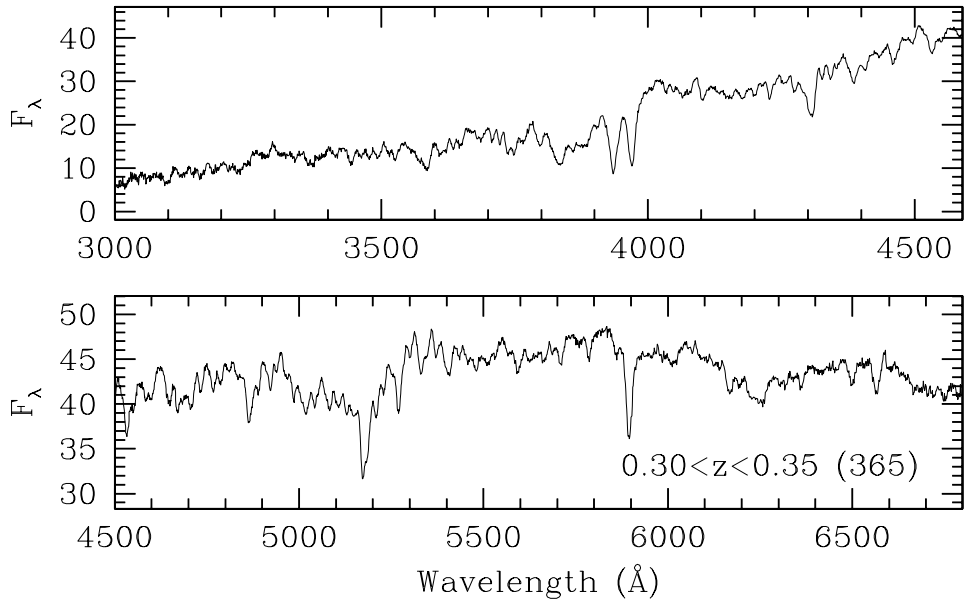}}
\caption{\label{fig:template}%
The average spectrum of 365 luminous red galaxies, selected as 
described in \S~\protect\ref{sec:lrg}, in the redshift
range $0.3<z<0.35$ over the rest-frame wavelength range of 3000\AA\
to 6800\AA.  The noise and pixel scale are visible on the blue and 
red ends as a fine-scale roughness; in between, the noise per pixel
is generally smaller than the thickness of the line.  
The vertical normalization is arbitrary.
}
\end{figure*}

\begin{figure*}[p]
\centerline{\epsfysize=7in \epsffile{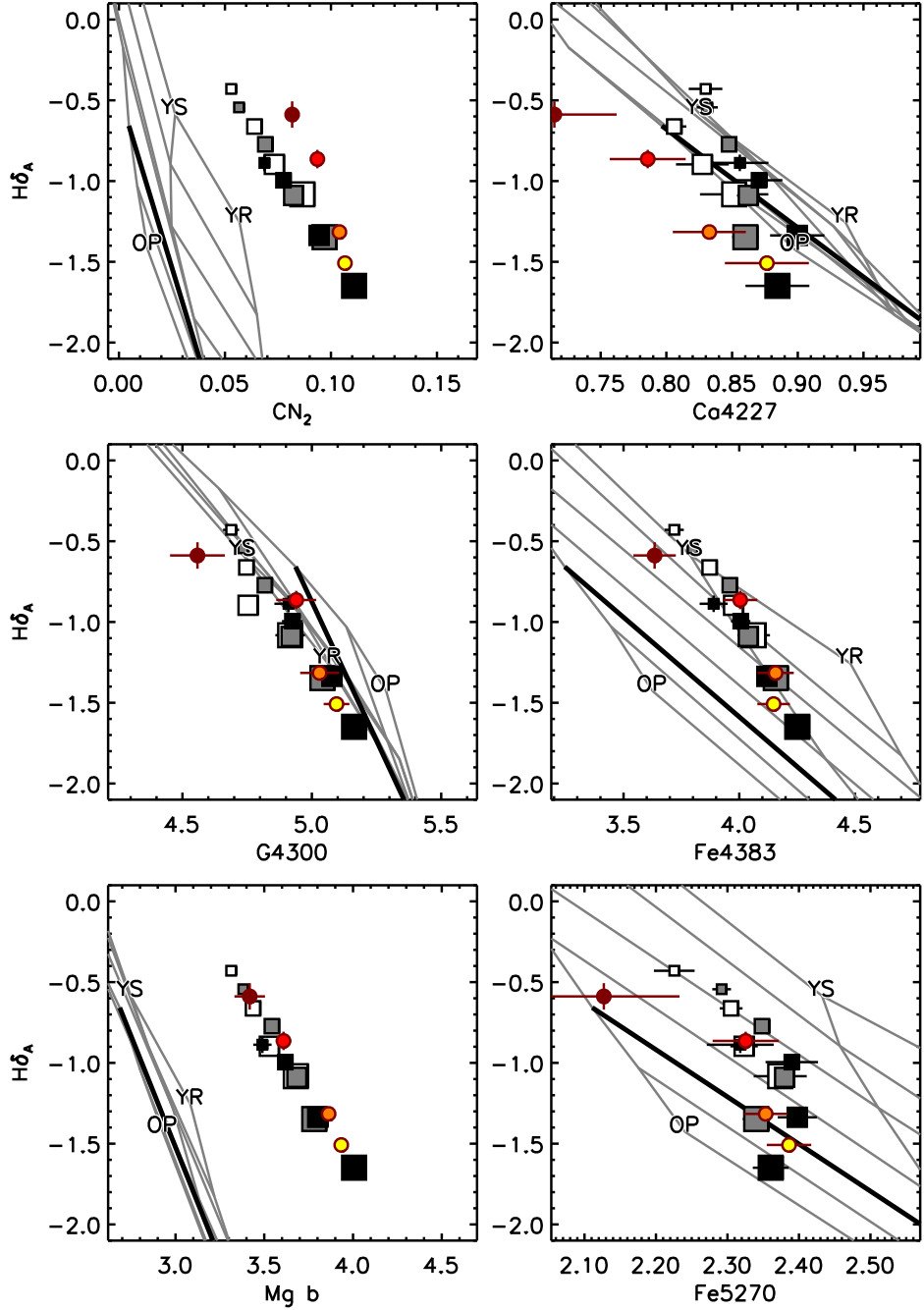}}
\caption{\label{fig:indexgrid}%
Each panel shows a plot of two sLick indices, with the H$\delta_A$ index
plotted against various metal indices.
Points show the values measured
from different samples.  Squares indicate the environment sub-samples.
Black squares are the $N>10$, high-density sample.  White squares 
are the $N\le3$, low-density sample.  Grey squares are the $4\le N\le9$,
intermediate-density sample.  The size of the square indicates the 
luminosity of the sample, with the largest square being $-22.5<M_r<-22$
and the smallest square being $-21<M_r<-20.5$.  Circles indicate
the LRG sample divided by redshift, with the darker color being higher redshift.
Error bars in both indices are shown but are often smaller than 
the points.  The line grid shows the predictions from the V99 model,
as described in \S~\protect\ref{sec:vazdekis}.  Three metallicities
are shown ($-0.4$, $0$, and $+0.2$~dex relative to solar) and six
different ages (4, 5, 6.3, 8, 10, 12, and 15~Gyr).  The thicker line marks
the 10 Gyr locus.  The letters YS, YR, and OP indicate the `young, solar',
`young, metal-rich', and `old, metal-poor' boundaries of the grid.
}
\end{figure*}

Figure \ref{fig:template} gives an example of the type of average spectrum
we are analyzing.  365 spectra of luminous early-type galaxies 
in the redshift range $0.30<z<0.35$ have been used to produce this
stacked spectrum.  The wavelength range of 3000\AA\ to 6800\AA\ rest-frame
is shown; this corresponds roughly to 4000\AA\ to 9000\AA\ in observed
wavelength.  Careful 
inspection of the blue and red ends reveals some noise on the pixel scale.
In the blue, this is caused by the lower sensitivity of the spectrographs
and of course the intrinsic faintness of early-type galaxies in the
rest-frame UV.  In the red, imperfect sky subtraction limits the 
performance.
In between, the error per pixel is smaller than the
thickness of the line.  
The velocity dispersion of the ensemble creates a visually obvious 
scale of the width of astrophysical features in the spectrum; 
comparing that width to the pixel scale revealed by the noise should
convince the viewer that essentially all of the features in the spectrum
are real.  Nearly all of the average spectra we consider have higher 
signal-to-noise ratios than this one.

\subsection{Absorption line strengths}

All of the average spectra in this paper are typical early-type 
galaxy spectra; they are
dominated by the absorption line features found in the spectra of old
stellar populations.  This can be made quantitative with comparision
to the V99 models via the sLick indices.

Broadly, the sLick index measurements (given in
Table~\ref{tab:slickellip1}) are consistent with the conclusion that 
our sample is composed primarily of an old,
metal-rich stellar population.
However, there is a large scatter in the inferred ages or
metallicities among different sLick indices; i.e., none of the V99
models is a good match to the total spectrum.  

Figure~\ref{fig:indexgrid} shows this graphically.  
In each panel, a sLick index based on \Hdelta\ is plotted 
against one of various metal line indices.
The measurements from 16 average spectra, our luminosity and environment
samples, are shown as square dots.
The V99 models are shown as a grid of metallicity and age.
Three metallicities ($-0.4$, 0, and $+0.2$ dex relative to solar)
and six ages (4, 5, 6.3, 8, 10, 12, and 15 Gyr) are shown.
We will defer discussion of the trends in luminosity and
environment until \S~\ref{sec:spectralvar}; here our point
is simply that no single V99 model matches all indices for a given spectrum.
$\alpha$-elements (e.g., Mg~b) tend to show higher metallicities, 
while iron-peak elements (e.g., Fe5270) show lower metallicities, 
This can be explained,
at least in part, by a mismatch in chemical abundances.  In
particular, the Vazdekis models have solar abundance ratios and it appears
that the average spectra show enhanced abundances for $\alpha$-process
elements relative to iron-peak elements.  This $\alpha$-enhancement has
been noted previously \citep{Wor92,Vaz99,Dav01,Tra00a,Pro02} and is expected for a
stellar population that converts its gas entirely into stars rapidly,
before type Ia supernovae have time to significantly pollute the
interstellar medium with iron-peak elements.

The hydrogen Balmer lines have been widely used to try to disentangle
metallicity and age estimation in old stellar systems
\citep[e.g.,][]{OCo76,Ros85,Wor94,Wor97,Vaz99b}.
We study the \Hbeta, \Hgamma, and \Hdelta\ lines.
For galaxies with velocity dispersions of $275\;\kms$, the Vazdekis
models predict that \Hbeta\ sLick index is the least sensitive to metal content,
while the \Hgamma\ and \Hdelta\ sLick indices still carry significant
metal dependence (in the sense that more metal-rich systems would
be inferred to be younger).  However, as we will see, \Hbeta\ 
has some contamination from interstellar emission lines and so we
focus on \Hdelta\ and \Hgamma\ indices, where the effect is negligible.  

For \Hdelta\ and \Hgamma, comparing the V99 models at solar metallicity
suggests a relatively young age, roughly 5--6 Gyr, for a single-age burst
(see the $H\delta_A$ versus Fe4383 panel of Figure \ref{fig:indexgrid}).  
All four indices ($H\delta_A$, $H\delta_F$, $H\gamma_A$, and $H\gamma_F$) 
give similar ages.  
As usual, a higher metallicity would imply a younger age.  
Such age inferences are famously biased by the presence of young A and F stars.
We therefore consider histories in which some fraction of
the stars are indistinguishably old (we adopt 12.5 Gyr) 
while the rest are formed at a constant
rate between 1 and 12.5 Gyrs (we exclude recent formation on the grounds
that our color cut would exclude blue star forming galaxies), assuming
solar metallicity throughout \citep[a poor assumption, see][]{Mar00}.  
To match \Hgamma\ and \Hdelta, one 
would need to form $\sim\!50$\% of the stars in the continuous phase, which
gives an average age of $\sim\!9$ Gyr.

$H\beta$ on the other hand implies an old age, above 12 Gyr for the
solar-metallicity V99 model.  The emission in the
average spectrum (\S~\ref{sec:emline}) undoubtably makes this an upper limit.  
When correcting for non-zero velocity dispersions 
and line emission by the prescriptions of 
\citet{Tra98,Tra00a}, we find $H\beta$, Mg~$b$, Fe5270, and Fe5335 indices
that fall comfortably within the $z=0$ locus found by \citet{Tra00a}, 
although our 
$H\beta$ lines are slightly deeper (i.e., younger) than their median value.

Above all, however, we feel that the obvious failings of solar 
abundance ratio models to simultaneously fit multiple metal indices 
caution against any
quantitative conclusion on the age of the stellar populations.  
The Balmer line indices are affected by metal absorption, 
and it may be that neglecting the $\alpha$ to Fe abundance ratio
differences skews the age inferences or disturbs the 
agreement between \Hdelta\ and \Hbeta.

\subsection{Emission-line components}
\label{sec:emline}

\begin{figure}[tb]
\noindent\centerline{\epsfxsize=\columnwidth\epsffile{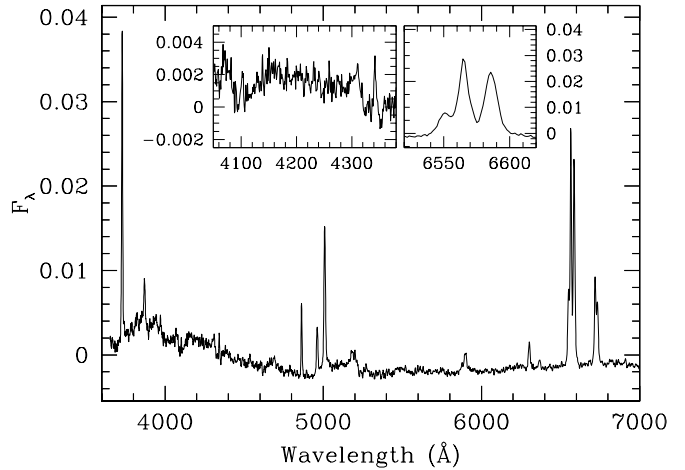}}
\caption{\label{fig:emline}%
The difference spectrum of outliers from the locus
of luminous bulge-dominated galaxies 
in the luminosity range $-22.5<M_r<-22$.
5\% of the galaxies whose PCA coefficients fall outside 
of the primary distribution of points are labeled as
outliers.  This spectrum is the change to the average
spectrum from including or excluding these objects.
The resulting spectrum shows a number of emission lines, with relative
strengths characteristic of a LINER, as well as some alterations
to some absorption lines, including the higher-order Balmer lines.
}
\end{figure}

In addition to generating the average spectrum, the PCA algorithm provides
us with all of the spectral components that contribute significantly
to the diversity (variance) of the spectra in our samples.  There is a
clear emission-line component that affects roughly 10\% of the
spectra.  We isolate all variations from the ``typical'' spectrum
by seeking outliers in the locus of PCA coefficients.  
In particular, noting that the PCA coefficients
of most galaxies fall in a tight distribution near zero (save for the
first component, which has a tight distribution with non-zero mean),
we form the sample of galaxies that have all coefficients (excluding
the first and as usual any beyond the 20th component) 
within some tolerance of zero.  The numerical value of this threshold
is meaningless without discussion of the normalization of the PCA
eigenspectra, but visually it imposes a tight envelope around the 
cloud of points containing the bulk of the galaxies.  It is likely that 
much of the scatter in this cloud is observational scatter.
We call the selected galaxies the ``median'' sample.  We then form
the average spectrum of this sample and subtract that from the average
spectrum of the full sample.  

An example of the resulting spectra is shown in Figure \ref{fig:emline}.
Here we have taken our MAIN sample bulge-dominated galaxies in the highest
luminosity bin, $-22.5<M_r<-22$.  Only 5\% of the galaxies fell in the
outlier sample, although this fraction rises at lower luminosities
(see Table \ref{tab:emline1}).
The residual spectrum has obvious emission lines, as well as reduced 
absorption at Na~D and Mg~$b$.
There is a hint of additional absorption in the higher Balmer lines
as well as an overall blue tilt.
It is worth noting, however, that the differences between the average
spectrum of all galaxies and the ``median'' sample are small, 
as one can see by comparing the vertical axes
in Figures \ref{fig:emline} and \ref{fig:averagespect}.
The effects on the sLick indices are negligible, save for the 
stronger emission lines (e.g. \OIII\ and \Hbeta).

We do not claim that this residual spectrum contains all of the line emission;
clearly a baseline emission rate will not be included.
However, if emission is the exception rather than the rule for these
galaxies, then this residual spectrum will contain most of the emission.
It is our impression that a few objects dominate the total emission strength.
Note that the method will select all the objects with strong \Halpha,
\OII, and/or \OIII\ emission; the weak emission at \Hgamma\ and \Hdelta\ 
need not be detected in individual spectra for it to appear 
in the residual spectrum.
Finally, it is important to note that the residual spectrum contains
all deviations from the median, but it is not required that these
deviations occur along a single ``direction'' in the space of PCA
coefficients.  While we detect emission and absorption, it is possible
that different outlier galaxies contribute these in different proportions.

In detail, the emission-line spectrum satisfies most of the definitions of
a LINER \citep{Hec80,Vei87}.  \NII\ $\lambda6583$, 
\OI\ $\lambda6300$, and \SII\ $\lambda\lambda6716,6731$ 
are strong compared to \Halpha, while 
\OIII\ $\lambda5007$ is weak compared to \Hbeta.
Of course, it is not 
straight-forward to give a physical interpretation to this average
spectrum.  It is likely that the galaxies show a diversity of star formation
and AGN emission mechanisms, and we would not claim that the
average spectrum proves that star formation is negligible.
Future work will explore the details of the emission line behavior
of these early-type galaxies.

Lower luminosity bulge-dominated galaxies show stronger emission-line
components as well as a more obvious Balmer absorption component. 
This will be discussed in \S~\ref{sec:emlinevar}.

\section{Variations with Luminosity and Environment}
\label{sec:spectralvar}

We now will proceed to our primary task, to analyze the average
spectra of luminous bulge-dominated galaxies as a function of luminosity 
and environment.
Recall that, as described in \S \ref{sec:select}, we break each of our
4 luminosity bins into 3 environment bins.  We also analyze each of
the 4 luminosity bins in its entirety.

There is a striking level of similarity among the average spectra of
these subsamples.  
Dividing by redshift, luminosity, and environment always yields the same
qualitative spectrum.  
However, we do find small but
highly significant differences among the average spectra.
In broad terms, the trends are similar to those found in previous work
\citep[e.g.,][]{Dre87,Bow90,Guz91,Ben93,Ros94,Tra00b,Ber01}.

\subsection{Variations in the sLick indices}

\begin{figure*}[p]
\centerline{\epsfysize=7in \epsffile{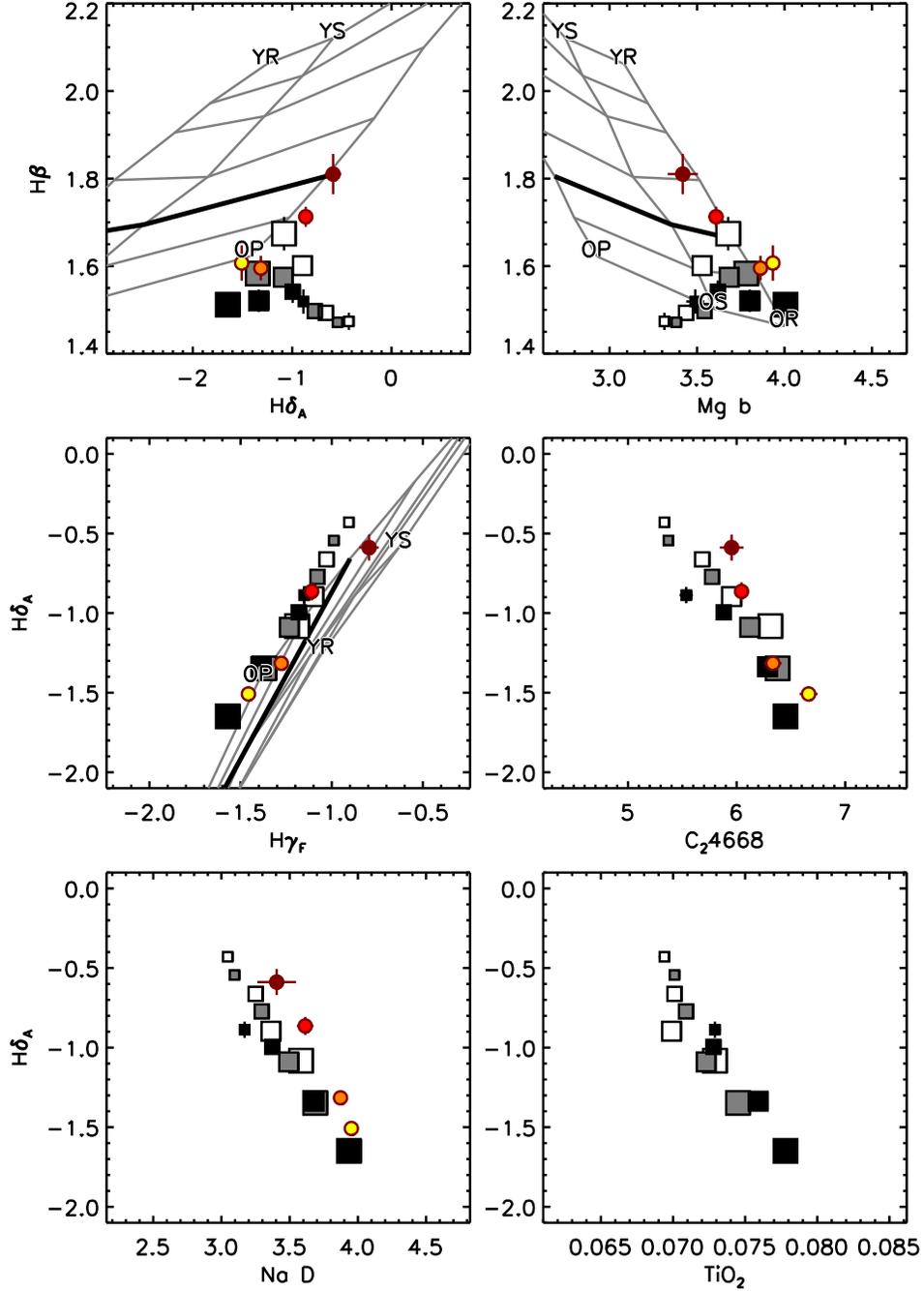}}
\caption{\label{fig:indexgrid1}%
As Figure \protect\ref{fig:indexgrid}, 
each panel shows a plot of two sLick indices.
The top two panels show the \Hbeta\ index against the 
\Hdelta$_A$ and $Mg_b$ indices.
These show that \Hbeta\ does not follow a simple one-dimensional
locus; it is unclear whether this is the result of emission line
contamination.
The lower four panels show the H$\delta_A$ index against 
the H$\gamma_F$ index and three metal-line indices
plotted against the H$\delta_A$ index.  The grid of V99
models are absent in three panels where the index wavelength falls outside the wavelength
range of the models.
}
\end{figure*}

Figures \ref{fig:indexgrid} and \ref{fig:indexgrid1}
show the spectral variations
as a function of luminosity and environment as measured by the sLick indices.
Lower luminosity galaxies or those in poorer environments have
indices that would be interpreted as younger or more metal-poor.
In some cases, notably the iron indices, the variations would
be interpreted primarily as age.  However, we are concerned
that the poor match between the data and the models in other indices 
(e.g.\ Mg~$b$) may cast doubt on such detailed interpretation.

The quantitative values of the sLick indices are presented in Table
\ref{tab:slickellip1}, along with their errors, 
and the V99 model predictions
(where available).
We also present the slopes of linear regressions versus luminosity,
environment, and redshift within the sets of comparison samples.
Slopes that are significant at 3--$\sigma$ are shown in boldface.

\subsection{A one-dimensional spectral space}
\label{sec:onedim}

\begin{figure*}[p]
\centerline{\epsfysize=7in\epsffile{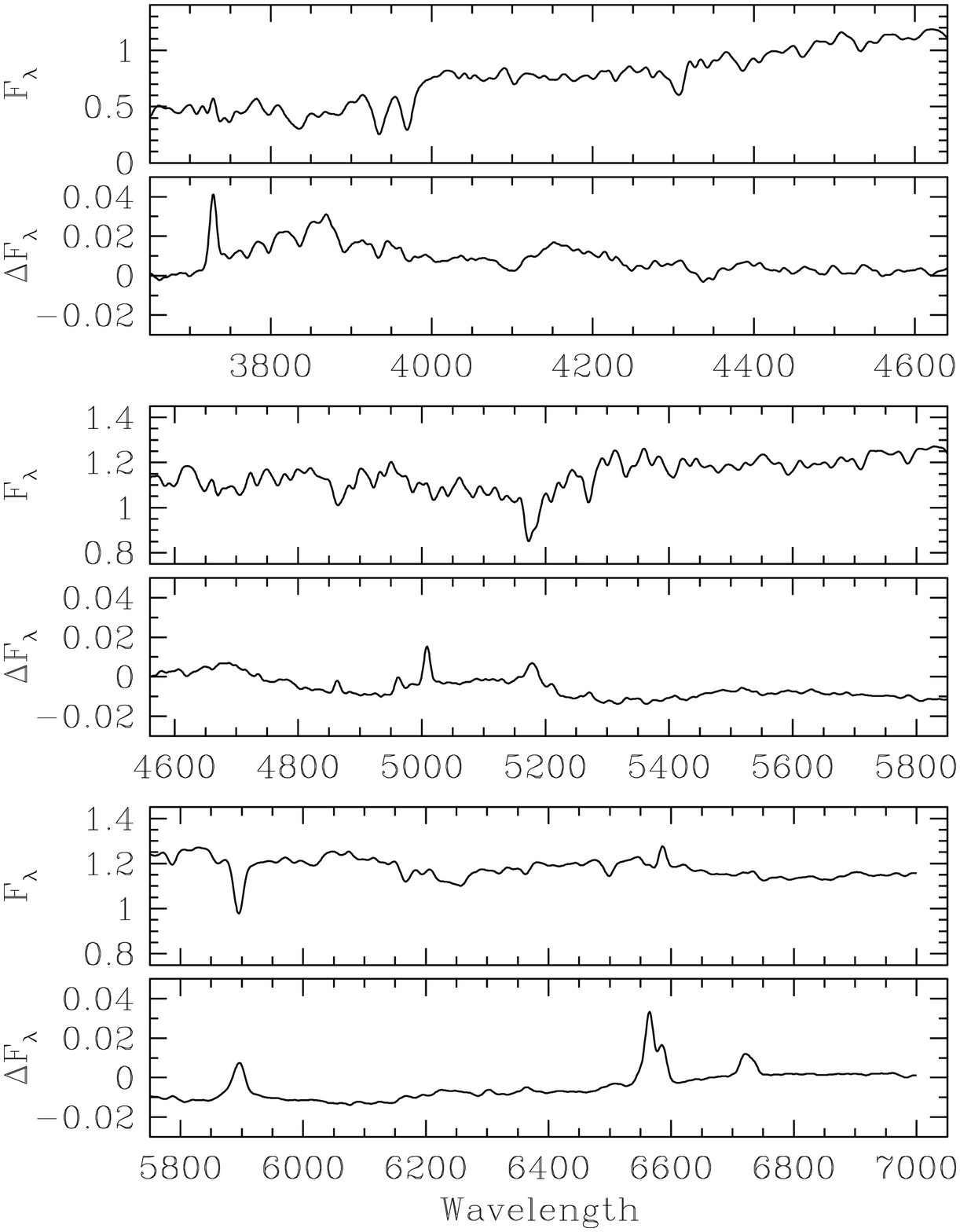}}
\caption{\label{fig:averagespect}%
The average of the average spectra from all of the environmental and
luminosity samples (note that this double counts some galaxies),
along with the first component of variation between these 16 sets.
This component accounts for 89\% of the variance in the 16 spectra.
The normalization of the first component of variation relative to the mean
spectrum is set by the
{\it rms} amplitude of this component.  Higher components have 
very small amplitudes ($\sim\!0.002$) on this scale.
}
\end{figure*}

\begin{figure}[tb]
\noindent\centerline{\epsfxsize=\columnwidth\epsffile{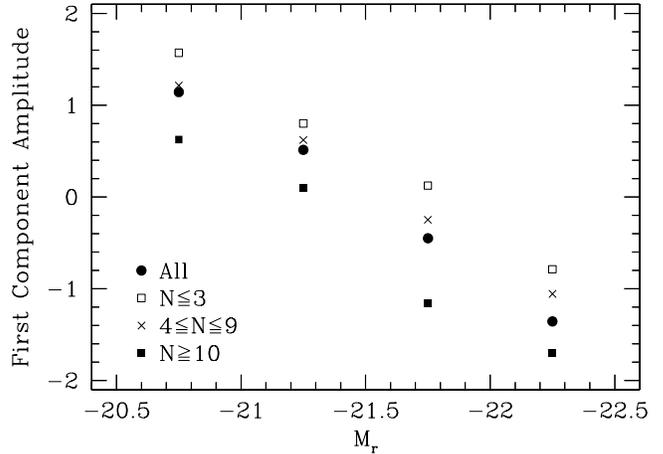}}
\caption{\label{fig:avervariations}%
The amplitude of the first component of variation relative to the
mean (these two spectra are shown in Fig.~\protect\ref{fig:averagespect})
as a function of luminosity for three different environment bins.
The normalization of the spectral component was chosen so that 
the {\it rms} value of this amplitude was unity; as shown in 
Fig.~\protect\ref{fig:averagespect}, this corresponds to $\sim\!1\%$
variations in the spectra.
}
\end{figure}

A striking feature of Figures \ref{fig:indexgrid} and \ref{fig:indexgrid1}
is that the locus of
values found in the luminosity and environment samples are nearly
one-dimensional.  In other words, with two different parameters at
play, one might have expected the sLick indices to populate
two-dimensional regions in these plots.  Instead, the points fall
(primarily) on tight curves, even for pairs of indices for which the
models predict a separation of the effects of age and metallicity.  
This suggests that luminosity and
environment control the same parameter in the average spectra.
The \Hbeta\ index is an exception, which we will discuss later.

To investigate this regularity, we took the 16 average spectra (12 from the 
luminosity/environment bins and 4 from the pure luminosity bins),
subtracted the mean, and performed a PCA analysis on the residuals.  
No points were masked and no continua were subtracted, but as usual the 
spectra were smoothed to $325\;\kms$ before performing the PCA.
89\% of the variance in the 16 spectra was contained in the first PCA
component, i.e., removing this component drops the rms variation from 1\%
to 0.3\%.  The mean spectrum and the first component of variation
is shown in Figure \ref{fig:averagespect}.  The normalization
of the first variation component relative to the mean is chosen so that
the {\it rms} amplitude of the variation is unity.  In other words,
the relative normalization in Figure \ref{fig:averagespect}
is the typical spread in the 16 spectra, which is $1\%$ {\it rms} of the mean.

The first component of variation is clearly 
a mix of emission and absorption lines.  
Emission at \Halpha, \Hbeta, \OII, \OIII, \NII, and \SII\ is visible.
Higher-order Balmer lines are seen in absorption, and the strong
Mg and Na absorption lines at 5175\AA\ and 5900\AA\ are varying in
strength.  The variation in the Mg line is slightly displaced to the 
red relative to the deepest part of the line in the mean spectrum;
this indicates that blending of the lines can be important.
Finally, there is a broadband tilt to the spectrum, in the sense
that variations in the direction of extra nebular emission and stronger
Balmer absorption correlate with a bluer spectrum, as one might expect.

The behavior at \Hdelta\ and \Hgamma\ is of concern:
although the variation spectrum appears to have extra absorption
at these freqencies, the absorption troughs are wider than expected
(compare the line shapes to those of the higher-order Balmer lines 
near 3800\AA).
In other words, we worry that other nearby metal lines may be impacting
the \Hdelta\ and \Hgamma\ sLick indices.  

The second component of variation in the PCA analysis 
is smaller by a factor of 4.  It is primarily
broadband fluctuations, with a small amount of \Halpha\ emission.

Figure \ref{fig:avervariations} shows the amplitudes of the first
component required by our 16 different samples.  As expected, 
lower luminosity galaxies and those in poorer environments have
more emission and Balmer absorption.

We stress that the information in Figure \ref{fig:averagespect} goes
well beyond the sLick indices.  We have found the 
empirical primary direction of variation with respect to luminosity
and environment for the ensemble of luminous, early-type galaxies
over most of the optical range.  While we are limited in the
resolution available from hot stellar systems, there are hundreds
of spectral elements measured to high signal-to-noise ratio in 
this spectrum.  Matching these variations would be a stringent
test for stellar population synthesis.  We have not attempted
more detailed comparisons to the models because it is clear that
the available solar element abundance ratio models 
are a poor match to the mean spectrum.

As an aside, the fact that the first component of spectral variation contains
89\% of the variance between the 16 spectra while being only 1\% variation
{\it rms} is a striking confirmation of the high signal-to-noise ratio
of the average spectra.  If these spectra had S/N of only 100 per pixel,
then the PCA eigenspectra corresponding to this noise would have
been comparable to the astrophysical variation, so that the
first component would have only explained $\sim\!1/16$ of the variance.
Hence, the signal-to-noise ratio of the spectra must considerably
exceed 100 for errors that are not common to all the spectra.
This is particularly remarkable given that no broadband modes were
removed prior to this PCA analysis; indeed, much of the first component's
variation is broadband, as it
contains only 0.5\% {\it rms} of variation when 4 broadband modes are removed.

\subsection{Variations in emission-line components}
\label{sec:emlinevar}

\begin{table*}[tb]\footnotesize
\addtocounter{table}{1}   %%% We want to make this table 3
\caption{\label{tab:emline1}}
\begin{center}
{\sc Emission and Absorption Line Fluxes \\}
%%% \begin{tabular}{cc cccccccc} 
\begin{tabular}{ccccccccccccccc}
\tableskip\tableline\tableline\tableskip
Magnitude & Environment & $F_{em}$ & \multicolumn{8}{c}{Emission} & \multicolumn{4}{c}{Absorption}\\
& & & \Halpha & \Hbeta & \Hgamma & \Hdelta & \SII & \NII & \OIII & \OII & \Halpha & \Hbeta & \Hgamma & \Hdelta \\
\tableskip\tableline\tableline\tableskip
$-22.5<M_r<-22$ & All &  5\% & 0.31 & 0.073 & 0.028 & 0.017 & 0.21 & 0.40 & 0.17 & 0.28 & \nodata & 0.005 & 0.065 & 0.074 \\
 & $N\ge 10$ &  7\% & 0.17 & 0.053 & 0.017 & 0.010 & 0.15 & 0.32 & 0.11 & 0.19 & \nodata & 0.005 & 0.019 & 0.012 \\
 & $4\le N\le 9$ &  7\% & 0.34 & 0.086 & 0.033 & 0.017 & 0.22 & 0.40 & 0.11 & 0.27 & \nodata & 0.052 & 0.098 & 0.106 \\
 & $0\le N\le 3$ &  6\% & 0.22 & 0.067 & 0.024 & 0.010 & 0.16 & 0.35 & 0.35 & 0.25 & \nodata & \nodata & 0.039 & 0.293 \\
\tableskip\tableline\tableskip
$-22<M_r<-21.5$ & All &  8\% & 0.64 & 0.133 & 0.047 & 0.026 & 0.33 & 0.62 & 0.33 & 0.36 & \nodata & 0.022 & 0.137 & 0.128 \\
 & $N\ge 10$ &  7\% & 0.37 & 0.073 & 0.021 & 0.015 & 0.20 & 0.30 & 0.13 & 0.20 & \nodata & 0.039 & 0.120 & 0.117 \\
 & $4\le N\le 9$ &  7\% & 0.62 & 0.124 & 0.049 & 0.031 & 0.33 & 0.57 & 0.39 & 0.39 & \nodata & 0.009 & 0.133 & 0.140 \\
 & $0\le N\le 3$ &  8\% & 0.65 & 0.125 & 0.048 & 0.019 & 0.30 & 0.58 & 0.26 & 0.32 & 0.061 & 0.032 & 0.187 & 0.139 \\
\tableskip\tableline\tableskip
$-21.5<M_r<-21$ & All & 10\% & 1.02 & 0.207 & 0.073 & 0.035 & 0.42 & 0.86 & 0.41 & 0.41 & \nodata & 0.052 & 0.239 & 0.225 \\
 & $N\ge 10$ & 11\% & 0.73 & 0.143 & 0.060 & 0.030 & 0.33 & 0.66 & 0.33 & 0.39 & \nodata & \nodata & 0.158 & 0.153 \\
 & $4\le N\le 9$ & 10\% & 0.82 & 0.172 & 0.062 & 0.022 & 0.35 & 0.68 & 0.27 & 0.35 & \nodata & 0.045 & 0.254 & 0.198 \\
 & $0\le N\le 3$ & 11\% & 1.10 & 0.216 & 0.078 & 0.037 & 0.43 & 0.92 & 0.43 & 0.41 & \nodata & 0.069 & 0.255 & 0.246 \\
\tableskip\tableline\tableskip
$-21<M_r<-20.5$ & All & 12\% & 1.34 & 0.289 & 0.106 & 0.052 & 0.53 & 1.03 & 0.58 & 0.51 & \nodata & 0.086 & 0.350 & 0.334 \\
 & $N\ge 10$ & 10\% & 0.85 & 0.185 & 0.078 & 0.046 & 0.43 & 0.79 & 0.51 & 0.49 & \nodata & 0.009 & 0.180 & 0.173 \\
 & $4\le N\le 9$ & 22\% & 1.26 & 0.270 & 0.096 & 0.046 & 0.47 & 0.96 & 0.40 & 0.42 & \nodata & 0.114 & 0.328 & 0.294 \\
 & $0\le N\le 3$ & 20\% & 1.26 & 0.264 & 0.097 & 0.045 & 0.50 & 0.93 & 0.47 & 0.46 & \nodata & 0.076 & 0.343 & 0.325 \\
\tableskip\tableline\tableskip
\multicolumn{3}{c}{Continuum $F_\lambda$, units $\AA^{-1}$} & 1.20 & 1.14 & 0.92 & 0.75 & 1.15 & 1.2 & 1.10 & 0.46 & 1.20 & 1.14 & 0.92 & 0.75 \\
\tableskip\tableline\tableline\tableskip
\end{tabular}

%%% \end{tabular}
\end{center}
NOTES.---%
Emission line strengths computed from the residual spectrum formed by
the difference of the average spectrum and the ``median'' spectrum.
$F_{em}$ is the fraction of spectra that were declared as outliers and
therefore omitted from the ``median'' spectrum.
The higher-order Balmer lines
are also seen to be in absorption; we fit a broad absorption Gaussian and a
narrow emission Gaussian.  
The results are quoted as line fluxes in arbitrary units; the $F_\lambda$ of 
the continuum is also given (in units of flux per \AA).  Equivalent widths
can be found by dividing the line flux by the continuum $F_\lambda$.
\SII\ is a sum of the 6731\AA\ and 6716\AA\ lines, \NII\ is a sum of the
6548\AA\ and 6583\AA\ lines, \OIII\ is only the 5007\AA\ line, and \OII\ 
is the sum of the 3726\AA\ and 3729\AA\ lines.
\end{table*}

In \S~\ref{sec:emline}, we isolated the spectral variations of outlier
galaxies by using the PCA coefficients to identify the outliers and 
then considering the difference between the average spectrum of all 
the galaxies and the average spectrum of those in the ``median'' sample.  
We now consider 
these residual spectra as functions of luminosity and environment.  
We generally find that galaxies of lower luminosity and poorer environment
have stronger emission line components.

In the case of \Hgamma\ and \Hdelta, we clearly see small emission
lines sitting in deeper absorption lines (Figure \ref{fig:emline} inset), 
indicating that the residual
spectrum contains both Balmer absorption from young stars and 
Balmer emission from young stars and active nuclei.  
We stress again that this superposition is a property of the composite
of outliers and may reflect multiple populations of outliers from
the ``typical'' old stellar spectrum.

To quantify the strength of these lines,
we fit separate Gaussians
to the emission and absorption lines, assuming that all emission lines
in the spectrum have the same velocity width.  The resulting line fluxes
are shown in Table \ref{tab:emline1}.  Converting these to equivalent
widths shows that the contribution to the \Hgamma\ and \Hdelta\
indices are very small: 
a 15\% effect on the inferred age in the worst case.  
We therefore infer that the 
\Hgamma\ and \Hdelta\ indices are negligibly affected by emission
in the full sample.

\Hbeta\ on the other hand does seem to have significant emission,
roughly enough to cancel out what must be non-negligible absorption.
The two contributions are more difficult to disentangle, and it seems
possible that the age inferred from \Hbeta\ would be significantly
affected.  
The fact that $H\delta$ and \Hgamma\ are well-correlated with metal-line 
indices while \Hbeta\ has a lot of scatter may also be suggestive
of emission in \Hbeta. However, it might indicate that the higher lines 
have significant contamination from metallicity variations.

\Halpha\ has even stronger emission and the presence of
strong \NII\ makes it impossible to detect an absorption component.
The line ratio of \NII\ to \Halpha\ displays a increasing trend with
luminosity and density,
which may indicate an decreasing role of star formation in driving
the emission \citep{Vei87}.

In general, the Balmer decrement is 
larger than expected from conventional 
predictions of the recombination spectrum.  This may be
caused by dust obscuration of the emission regions.  

\subsection{Selection biases}

The interpretation of the spectral differences is subject to considerations
of biases in the sample selection and analysis.

Fortunately, the comparison between samples from different environments
is substantially immune to redshift-dependent effects, such
as variations in the sample selection or change in physical spectroscopic 
aperture across redshift,
because the redshift distributions of the three environment bins are
nearly identical.
Target selection and all aspects of the observations were done without
consideration of the environment of the galaxies.  The only
significant effect is that galaxies in clusters tend to be more
luminous.  However, the 0.5 magnitude bin size in luminosity limits
the shift in the mean luminosity between environment subsamples
to less than 2\% in the lower three luminosity samples 
and 4\% in the most luminous sample (Table \ref{tab:summary}).

The comparison across luminosity, on the other hand, may include some
biases in sample selection.  First, there is a small shift in redshift
(listed in Table~\ref{tab:summary} and shown in Figure~\ref{fig:redshift})  between
the luminosity bins.  If all samples were passively evolving, the most
luminous galaxies would appear $\sim 1$~Gyr younger than the least
luminous relative to a fixed redshift comparison.  Second, galaxies of
different luminosities have different physical sizes, so that the spectroscopic fiber
will include a different fraction of, e.g., the effective radius of
the de~Vaucouleurs model.  Radial population gradients will therefore
bias our results \citep[and references therein]{McC69,Mun92}.  
This effect is partially offset by the change in
angular diameter distance with redshift.  Finally, and most important,
the lower luminosity bins almost certainly have more contamination
from objects that would not be classified by eye as red, elliptical
galaxies.  Our morphology cut is relatively crude and does not include
obvious mainstays of quantitative morphology such as bulge-disk
decomposition and smoothness.  It is not clear that
equivalent samples are being selected at all luminosities or even
that equivalent samples can be selected in principle.
The difficulty of homogenizing stellar disk contributions 
could also bias the environmental subsamples
because of the density-morphology relation.

\subsection{Interpretation of absorption lines}

Figures \ref{fig:indexgrid}, 
\ref{fig:indexgrid1}, and \ref{fig:averagespect} 
and Table \ref{tab:slickellip1} all demonstrate
that there are statistically significant differences between
the luminosity and environment subsamples.  Table \ref{tab:slickellip1}
displays the slopes and errors resulting from least-squares fits
to each sLick index across luminosity and environment.  Indices
with more than 3-$\sigma$ detections of variation are shown in 
boldface.
In particular, the \Hdelta, CN$_2$, \Hgamma, G4300, C$_2$4668, Mg~$b$, 
Na D, and TiO$_2$ indices all show strong variations.

In principle, such well detected variations should be able not only to
determine age and mean metallicity shifts, but also to probe for variations
in abundance ratios.
\citet[hereafter TB95]{Tri95} calculate the variation in Lick indices  
induced by changes in the abundances of individual elements for 
3 specific model stars.  
This grid of abundances and stars is too sparse 
to interpret the observed differences in our data.  
Nevertheless, if one adopts the overly
simple model in which the TB95 cool giant and hot dwarf models are combined
in arbitrary ratio with abundance variations allowed in groups of
C \& N, O, $\alpha$ elements, and iron peak elements, and then resolves our 
observed index variations against these 5 vectors of model variations, then one 
finds that the variations amongst our spectra are best described by 
variations in the
$\alpha$ element abundance.  Of course, age variations are available
to this model only in so far as they could be mapped to changes in
the giant to dwarf ratio.
As an additional caveat (were there
not enough already!), we note that the TB95 Lick indices are quoted at zero
velocity dispersion while our variations are at $275\;\kms$.

The indication that abundance ratio variations may be the 
controlling parameter in the spectral variation with luminosity
and environment is intriguing.
We believe that this hint should offer strong encouragement for the
construction of spectral models of stellar populations with non-solar
abundance ratios \citep[e.g.][]{Tho02}.

\section{Average spectrum vs cosmic time}

\begin{figure*}[p]
\centerline{\epsfysize=7in\epsffile{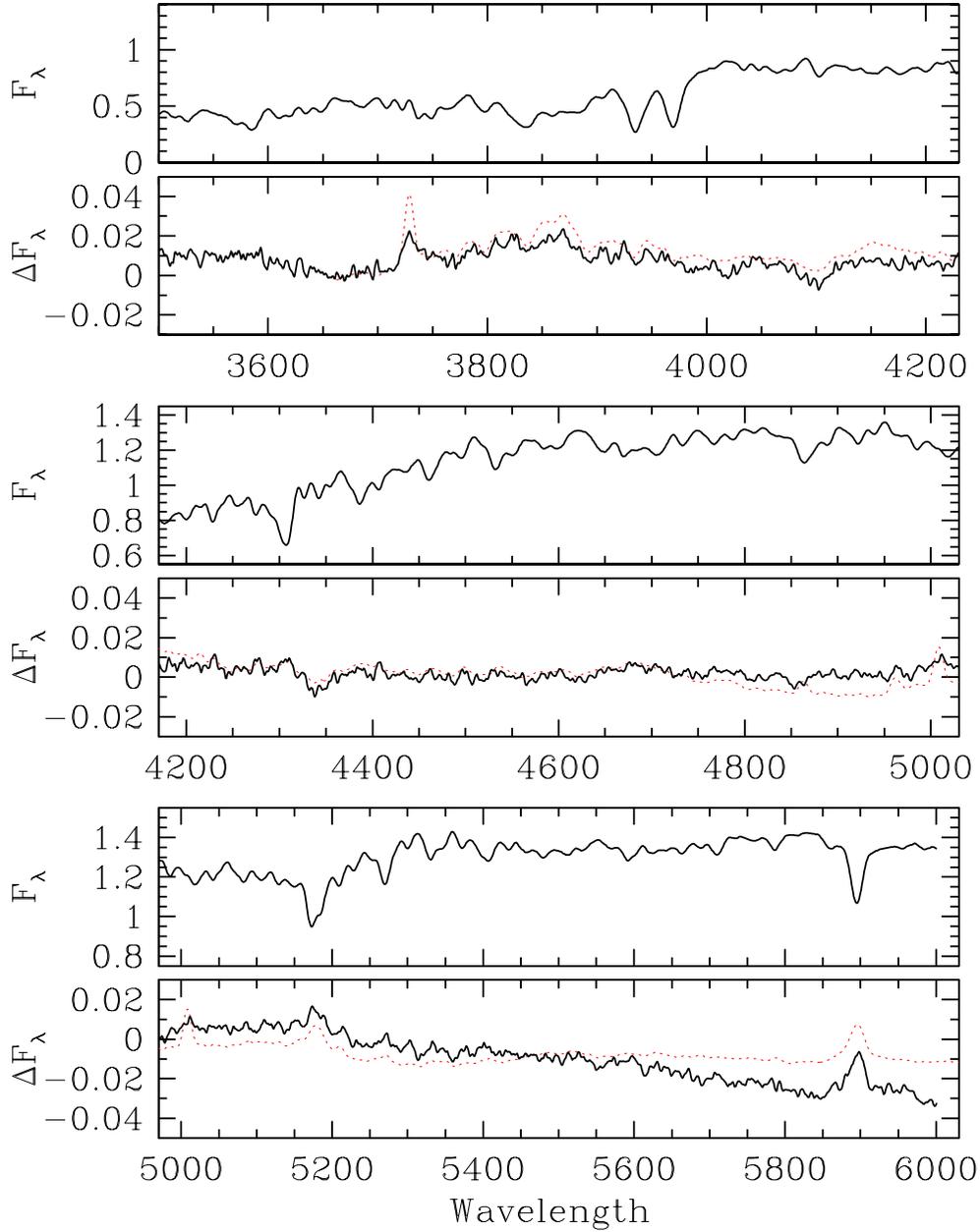}}
\caption{\label{fig:zspect}%
The average of the average spectra from all of the redshift
samples
along with the least-squares variation per 0.1 in redshift.
The sign of the variation is such that one should add the variation
spectrum to the mean when considering higher redshifts.
The first component variation spectrum from 
Figure \protect\ref{fig:averagespect} 
is shown for comparison ({\it dotted line}).  The
relative normalization of the two variation spectra is arbitrary.
The broadband offsets of the redshift variation spectrum should
not be trusted because of spectrophotometry uncertainties; comparing
fixed rest-frame wavelengths across redshift requires accurate 
spectrophotometry across a significant range of observed wavelengths, 
whereas the comparisons across environment and luminosity were less 
demanding of this.
The redshift variation spectrum has more small-scale noise in part because
the more luminous galaxies required less smoothing to reach a velocity
dispersion of $325\;\kms$, thereby including more photon noise.  
Simulation of the noise indicates that, for example,
the three largest features between 4225\AA\ and 4325\AA\ are
significant at roughly 5-$\sigma$ confidence, while the features
between 4150\AA\ and 4200\AA\ are only marginally significant.
}
\end{figure*}

Whereas our comparison of MAIN samples was all at low redshift,
our LRG sample spans a large range in redshift, $0.15 < z < 0.50$, 
with galaxies that are similar in luminosity and rest-frame color.
We can therefore look for the evolution of the average spectrum 
across cosmic time.  We divide the galaxies into 4 separate redshift bins:
$0.15 < z < 0.25$, $0.25 < z < 0.35$, $0.35 < z < 0.45$, and $0.45 < z < 0.50$.

While the selected galaxies are luminous early-types at all redshift,
there are subtle redshift dependences in the selection.  
For example, the fine details of
the rest-frame luminosity-color region imposed at each redshift by 
the LRG selection will have mild redshift dependences (Figure \ref{fig:lrgsel}).
At this point, we have not assessed these 
because of uncertainties in the appropriate $k$-corrections and
evolution corrections as well as the need to model the effect of the {\it noise} in
the observed colors and magnitudes on the sample selection.
Detailed modelling of these selection effects is now underway.

To first order, the galaxies in each bin should be similar,
in the sense that a passively-evolving old galaxy would be in all the
samples.  However, the physical aperture of our $3''$ diameter
spectroscopic fiber is increasing with redshift, so that we are
including more of the light of the higher redshift galaxies.  If galaxies
have older or more metal-rich stellar populations near their centers, 
then the higher redshift objects will be biased to look younger or
more metal-poor.  Of course, the size of such effects is bounded by
the weakness of color gradients in giant elliptical galaxies 
\citep{Fra89,Pel90,Mic00}
and by the fact that we observe the luminosity-weighted spectrum 
within the physical aperture.

The sLick indices measured for these average spectra are given in
Table \ref{tab:slickellip1} and plotted as circles in 
Figures \ref{fig:indexgrid} and \ref{fig:indexgrid1}.  
The results show a clear dependence on redshift.  Were this
difference due only to age, the higher redshift galaxies would
be younger by $\sim\!3$~Gyr, with some variation with index.
This is consistent with concordance cosmology prediction of $\sim2.5$~Gyr.
In the Figures, one should note that 
the results do not in all cases lie along the locus defined
by the low-redshift luminosity and environment variations.
If we view the redshift sample as defining the true time derivative
of a metal-rich passively evolving population, then the luminosity
and environment dependences may require abundance variations.

We next consider the variations of the full spectrum across redshift.
We take the four average spectra, subtract the mean, and compute
the least-squares slope for each pixel as a function of the central
redshift of each spectrum.  The vector of slopes is the spectrum of
ordered variations within this set of four spectra.  
The highest redshift average spectra has
noise roughly twice as large as the others, so we downweight it by a
factor of 4 in the least-squares fitting.  
We normalize the slope to a difference of 0.1 in redshift.  
The PCA analysis of \S~\ref{sec:onedim} yields a similar variation
spectrum to this least-squares method.

The resulting mean and variation spectra are shown in Figure
\ref{fig:zspect}.  The variation spectrum shows more noise than the 
equivalent spectrum in the environment/luminosity comparison.  This is
partially because the highest redshift average spectra are noisier, but
it is also because the LRG sample spectra required less smoothing to
match to $325\;\kms$ than the MAIN sample spectra.  Despite this, many
features in the variation spectrum are well detected.  We assess
the significance by creating realizations of the noise of the original
spectra, smoothing, and applying the least-squares analysis.  The 
smaller features in the variation spectrum are not significant, but
features such as the three peaks between 4225\AA\ and 4325\AA\ are
clearly significant.  Further SDSS data should make interesting
improvements in the signal-to-noise ratio of these spectra.

The variation spectrum shows strong alterations to absorption 
features at \Hdelta, \Hgamma, Mg~$b$, and Na~D, as well as weaker
changes to other lines.
Mg~$b$ and Na~D are shallower at higher redshift, while the two Balmer
lines are deeper.
Little hint of emission lines is found, save for \OII, 
probably in part because the galaxies in the LRG sample are 
considerably more luminous than those in the MAIN sample.  
Overlaying the component of variation from the environment and
luminosity analysis shows matching trends in the strongest features 
but imperfect agreement in the amplitudes
of variations in different lines.  Hence, the two variation 
spectra (in this section and \S~\ref{sec:onedim}) represent
two different combinations of age and metallicity; one might
speculate that the redshift variation is dominantly due to age.

\section{Ultraviolet average spectrum}

\begin{figure}[tb]
\noindent\centerline{\epsfxsize=\columnwidth\epsffile{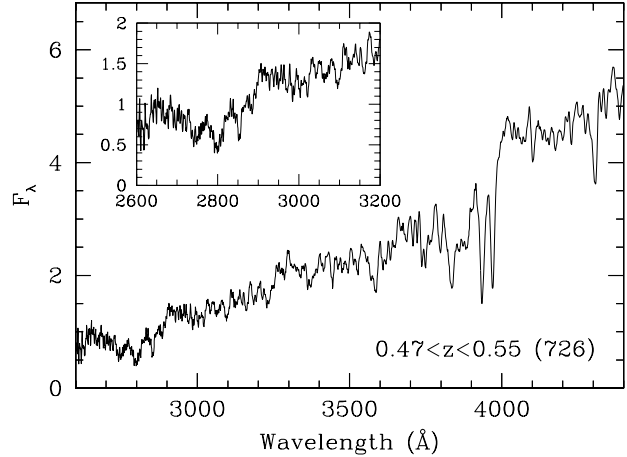}}
\caption{\label{fig:faruv}%
Average UV spectrum of 726 luminous, red galaxies at $0.47<z<0.55$.
The inset shows an expansion of the bluest region of the spectrum.
The 2900\AA\ and 2640\AA\ spectral breaks are clearly visible, as 
are the interstellar Mg lines at 2796\AA\ and 2852\AA.
The vertical normalization is arbitrary.
}
\end{figure}

In Figure \ref{fig:faruv}, we present the composite UV spectrum 
from our highest redshift galaxies, 726 luminous, red galaxies
at $0.47<z<0.55$. Many absorption lines are detected, although
the signal-to-noise ratio is poor at the bluest wavelengths
because of the dropping efficiency of the SDSS spectrographs
shortward of 4000\AA.  
The 2900\AA\ and 2640\AA\ spectral breaks are clearly visible,
as are the strong Mg lines at 2796\AA\ and 2852\AA.
Using the definitions of \citet{Spi97}, we measure the
strengths of these breaks to be
$B(2640) = 1.33\pm 0.06$ and $B(2900) = 1.41 \pm 0.04$.

The strengths of the UV breaks suggest that the UV spectrum has
contributions from F-type stars \citep{Spi97,Fio97}.  
Such stars can be present in the
spectrum because either (a)~there has been some star formation within
the previous few Gyr, or (b)~there is enough metallicity range among
stars in the composite spectrum that there are old, F-type stars,
either dwarfs or giants.
Both explanations are plausible; we expect some ongoing star formation
even among these old stellar populations, and we expect a wide
metallicity range.

\section{Summary and conclusions}

We have defined samples of luminous, red, bulge-domi\-nated galaxies
from the SDSS that represent controlled variations in luminosity,
environment, and redshift.  We have focused entirely on 
luminous galaxies, roughly $L^*$ and above.  The samples are
a nearly complete representation of the most massive galaxies in the local
universe.

We have computed luminosity-weighted average spectra so as to probe
the aggregate stellar populations of these galaxies.  The
spectra indicate fairly old, metal-rich stellar populations that are 
enhanced in $\alpha$ elements
relative to iron-peak elements.  Our primary goal has been to compare
the relative differences in the populations as a function of luminosity,
environment, and redshift.  We find that the spectral variations
across luminosity and environment are nearly a one-dimensional set.
With a velocity-broadened form of the Lick indices,
most index-index comparisons track a one-dimensional locus despite
having two independent parameters (luminosity and environment).
This is true even for pairs of indices that are predicted to split
the familiar age-metallicity degeneracy.
We find some indication that the controlling
parameter of the variation may be the $\alpha$-to-Fe abundance ratio.  
Our results include not simply a vector
of sLick index offsets, but a full resolution spectrum of the variation
that includes over a hundred clearly visible lines.

We compared the average spectra directly to models from V99, 
analyzing these model spectra
identically to our data, and find that the variations are plausible
even if the mean values of the indices may be shifted.  More
luminous galaxies, lower redshift galaxies, or those in denser
environments would be interpreted as more metal-rich or older. 
However, we stress that the obviously non-solar $\alpha$ to 
iron-peak ratio in our data probably invalidate any detailed interpretation
of the differences with the solar-abundance-ratio models of V99.

We hope that the detail available in these spectra and their derivatives
with luminosity, environment, and redshift will be a stimulus to 
detailed, full-resolution modelling of $\alpha$-enhanced, super-solar 
stellar populations.  The spectra are available in electronic
form as an on-line attachment to this paper.

\acknowledgements We thank Jim Peebles, Scott Trager, and Ann
Zabludoff for useful conversations and the referee for helpful comments.  
D.J.E. thanks NYU for hospitality
while a portion of this research was completed.  D.J.E. was supported
by National Science Foundation (NSF) grant AST-0098577, by Hubble
Fellowship grant \#HF-01118.01-99A from the Space Telescope Science
Institute, which is operated by the Association of Universities for
Research in Astronomy, Inc, under NASA contract NAS5-26555, and by a
Alfred P. Sloan Research Fellowship.  D.W.H. was partially supported
by NSF grant 0101738 and NASA contract NAG5-11669.  This
research made use of the NASA Astrophysics Data System.

Funding for the creation and distribution of the SDSS Archive has been
                    provided by the Alfred P. Sloan Foundation, the
                    Participating Institutions, the National
                    Aeronautics and Space Administration, the National
                    Science Foundation, the U.S.  Department of
                    Energy, the Japanese Monbukagakusho, and the Max
                    Planck Society. The SDSS Web site is
                    http://www.sdss.org/.

                    The SDSS is managed by the Astrophysical Research
                    Consortium (ARC) for the Participating
                    Institutions. The Participating Institutions are
                    The University of Chicago, Fermilab, the Institute
                    for Advanced Study, the Japan Participation Group,
                    The Johns Hopkins University, Los Alamos National
                    Laboratory, the Max-Planck-Institute for Astronomy
                    (MPIA), the Max-Planck-Institute for Astrophysics
                    (MPA), New Mexico State University, Princeton
                    University, the United States Naval Observatory,
                    and the University of Washington.

\begin{table*}[p]\footnotesize
\addtocounter{table}{-2}  %%% We want to make this Table 2
\caption{\label{tab:slickellip1}}
\begin{center}
{\sc sLick indices on Average Spectra\\}
%%% \begin{tabular}{cc cccccccc} 
\begin{tabular}{ccccccccc}
\tableskip\tableline\tableline\tableskip
Magnitude & Environment/  & H$\delta_A$ & H$\delta_F$ & CN${}_2$ & Ca4227 & H$\gamma_A$ & H$\gamma_F$ & G4300\\
& Redshift & \AA & \AA & mag & \AA & \AA & \AA & \AA\\
\tableskip\tableline\tableline\tableskip
$-22.5<M_r<-22.0$ & All
 & $-1.347$ (19) & $0.301$ (13) & $0.0970$ (5) & $0.860$ (10) & $-5.508$ (17) & $-1.371$ (11) & $5.041$ (22)\\
$-22.0<M_r<-21.5$ & All
 & $-1.089$ (14) & $0.437$ (10) & $0.0824$ (5) & $0.862$ (8) & $-5.224$ (12) & $-1.233$ (8) & $4.920$ (18)\\
$-21.5<M_r<-21.0$ & All
 & $-0.772$ (13) & $0.575$ (10) & $0.0690$ (5) & $0.848$ (6) & $-4.915$ (13) & $-1.079$ (8) & $4.818$ (14)\\
$-21.0<M_r<-20.5$ & All
 & $-0.544$ (19) & $0.650$ (14) & $0.0566$ (4) & $0.831$ (8) & $-4.729$ (17) & $-0.989$ (12) & $4.735$ (19)\\
\tableskip\tableline\tableskip
Slope per mag & 
 & {\boldmath $-0.552$} & {\boldmath $-0.242$} & {\bf +0.0267} & +0.021 & {\boldmath $-0.536$} & {\boldmath $-0.264$} & {\bf +0.203}\\
Error & 
 & $0.016$ & $0.012$ & $0.0004$ & $0.008$ & $0.015$ & $0.010$ & $0.018$\\
\tableskip\tableline\tableline\tableskip
$-22.5<M_r<-22.0$ & $N\ge10$
 & $-1.649$ (37) & $0.189$ (26) & $0.1109$ (13) & $0.884$ (24) & $-5.818$ (32) & $-1.571$ (20) & $5.163$ (39)\\
$-22.5<M_r<-22.0$ & $4\le N\le9$
 & $-1.259$ (23) & $0.334$ (16) & $0.0925$ (11) & $0.870$ (21) & $-5.447$ (29) & $-1.349$ (19) & $5.011$ (39)\\
$-22.5<M_r<-22.0$ & $0\le N\le3$
 & $-1.082$ (47) & $0.408$ (34) & $0.0865$ (11) & $0.852$ (26) & $-5.170$ (44) & $-1.191$ (29) & $4.916$ (57)\\
\tableskip\tableline\tableskip
$-22.5<M_r<-22.0$ & $N$ slope
 & {\boldmath $-0.599$} & {\boldmath $-0.231$} & {\bf +0.0236} & +0.033 & {\boldmath $-0.662$} & {\boldmath $-0.389$} & {\bf +0.256}\\
 & Error
 & $0.059$ & $0.042$ & $0.0017$ & $0.036$ & $0.054$ & $0.035$ & $0.068$\\
\tableskip\tableline\tableline\tableskip
$-22.0<M_r<-21.5$ & $N\ge10$
 & $-1.337$ (32) & $0.329$ (23) & $0.0944$ (11) & $0.900$ (20) & $-5.503$ (29) & $-1.389$ (20) & $5.078$ (40)\\
$-22.0<M_r<-21.5$ & $4\le N\le9$
 & $-1.069$ (24) & $0.424$ (18) & $0.0809$ (7) & $0.840$ (9) & $-5.264$ (20) & $-1.258$ (13) & $4.946$ (22)\\
$-22.0<M_r<-21.5$ & $0\le N\le3$
 & $-0.897$ (32) & $0.527$ (23) & $0.0733$ (11) & $0.827$ (20) & $-4.982$ (23) & $-1.099$ (15) & $4.754$ (31)\\
\tableskip\tableline\tableskip
$-22.0<M_r<-21.5$ & $N$ slope
 & {\boldmath $-0.440$} & {\boldmath $-0.198$} & {\bf +0.0208} & +0.071 & {\boldmath $-0.525$} & {\boldmath $-0.293$} & {\bf +0.333}\\
 & Error
 & $0.046$ & $0.033$ & $0.0016$ & $0.029$ & $0.038$ & $0.025$ & $0.051$\\
\tableskip\tableline\tableline\tableskip
$-21.5<M_r<-21.0$ & $N\ge10$
 & $-0.995$ (37) & $0.478$ (27) & $0.0776$ (11) & $0.871$ (17) & $-5.176$ (29) & $-1.181$ (18) & $4.925$ (40)\\
$-21.5<M_r<-21.0$ & $4\le N\le9$
 & $-0.818$ (22) & $0.546$ (15) & $0.0697$ (7) & $0.840$ (11) & $-4.962$ (21) & $-1.101$ (14) & $4.824$ (24)\\
$-21.5<M_r<-21.0$ & $0\le N\le3$
 & $-0.662$ (25) & $0.594$ (18) & $0.0639$ (5) & $0.806$ (9) & $-4.813$ (20) & $-1.029$ (14) & $4.747$ (26)\\
\tableskip\tableline\tableskip
$-21.5<M_r<-21.0$ & $N$ slope
 & {\boldmath $-0.328$} & {\boldmath $-0.112$} & {\bf +0.0132} & {\bf +0.065} & {\boldmath $-0.354$} & {\boldmath $-0.152$} & {\bf +0.173}\\
 & Error
 & $0.044$ & $0.032$ & $0.0012$ & $0.019$ & $0.035$ & $0.023$ & $0.046$\\
\tableskip\tableline\tableline\tableskip
$-21.0<M_r<-20.5$ & $N\ge10$
 & $-0.887$ (51) & $0.478$ (37) & $0.0686$ (14) & $0.856$ (21) & $-5.069$ (45) & $-1.152$ (29) & $4.910$ (54)\\
$-21.0<M_r<-20.5$ & $4\le N\le9$
 & $-0.585$ (28) & $0.625$ (20) & $0.0574$ (7) & $0.839$ (13) & $-4.758$ (23) & $-1.010$ (15) & $4.750$ (28)\\
$-21.0<M_r<-20.5$ & $0\le N\le3$
 & $-0.429$ (33) & $0.685$ (24) & $0.0529$ (9) & $0.830$ (12) & $-4.560$ (28) & $-0.906$ (19) & $4.688$ (31)\\
\tableskip\tableline\tableskip
$-21.0<M_r<-20.5$ & $N$ slope
 & {\boldmath $-0.427$} & {\boldmath $-0.190$} & {\bf +0.0141} & +0.024 & {\boldmath $-0.482$} & {\boldmath $-0.237$} & {\bf +0.194}\\
 & Error
 & $0.059$ & $0.043$ & $0.0016$ & $0.024$ & $0.051$ & $0.033$ & $0.058$\\
\tableskip\tableline\tableline\tableskip
$M_g<-21.8$ & $0.15<z<0.25$
 & $-1.508$ (47) & $0.228$ (34) & $0.1066$ (15) & $0.876$ (32) & $-5.590$ (43) & $-1.457$ (28) & $5.095$ (49)\\
$M_g<-21.8$ & $0.25<z<0.35$
 & $-1.315$ (41) & $0.301$ (30) & $0.1040$ (13) & $0.833$ (27) & $-5.332$ (56) & $-1.277$ (37) & $5.030$ (75)\\
$M_g<-21.8$ & $0.35<z<0.45$
 & $-0.864$ (56) & $0.587$ (42) & $0.0936$ (19) & $0.786$ (28) & $-4.978$ (51) & $-1.110$ (34) & $4.940$ (77)\\
$M_g<-21.8$ & $0.45<z<0.50$
 & $-0.588$ (81) & $0.644$ (60) & $0.0817$ (23) & $0.715$ (47) & $-4.315$ (78) & $-0.795$ (53) & $4.558$ (106)\\
\tableskip\tableline\tableskip
 & Slope per $\Delta z =0.1$
 & {\bf +0.319} & {\bf +0.158} & {\boldmath $-0.0080$} & {\boldmath $-0.051$} & {\bf +0.378} & {\bf +0.201} & {\boldmath $-0.138$}\\
 & Error
 & $0.027$ & $0.020$ & $0.0008$ & $0.017$ & $0.025$ & $0.017$ & $0.033$\\
\tableskip\tableline\tableline\tableskip
Model: 8 Gyr & [Z/H]=0.0
 & $-1.849$ & $0.074$ & $0.0353$ & $0.981$ & $-6.272$ & $-1.471$ & $5.338$\\
12 Gyr - 8 Gyr & [Z/H]=0.0
 & $-1.183$ & $-0.480$ & $0.0261$ & $0.216$ & $-1.179$ & $-0.562$ & $0.245$\\
8 Gyr & [Z/H]=0.2 - solar
 & $-0.941$ & $-0.296$ & $0.0502$ & $0.125$ & $-0.906$ & $-0.362$ & $0.194$\\
\tableskip\tableline\tableline\tableskip
\end{tabular}

%%% \end{tabular}
\end{center}
NOTES.---%
Stacked spectra and models \protect\citep{Vaz99} are smoothed until the
measured velocity dispersion is $\sigma=325\;\kms$.  The smoothed
spectra are then further convolved with a Gaussian to achieve the
resolution that a $\sigma=275\;\kms$ galaxy would have with the Lick/IDS
system.  However, unlike the official Lick system, we have kept the
spectra in their approximately flux-calibrated form.  No attempt
to correct the indices to zero velocity-dispersion has been made.
The focus is on comparison between the different samples.
The 1-$\sigma$ error in each measurement is given in parentheses.

The top set compares different luminosities within the MAIN set.  
The middle sets compare different environments within a given luminosity bin.
The final set compares different redshifts within the LRG set.
Each set is fit to a linear regression, and the slope and its error
are quoted.
The slopes versus environment are computed from the best-fit line considering the
three bins to be given abscissa values of 0, 0.5, and 1 from least dense
to most dense.
Boldface numbers indicate 3-$\sigma$ detections of a non-zero slope.

Model spectra from \protect\citep{Vaz99} are measured in the same
fashion.  Results for the 8 Gyr, solar metallicity model are quoted.
Also quoted are the difference between the 12 Gyr and 8 Gyr model,
both at solar metallicity, and the $[{\rm Z}/{\rm H}]=0.2$ and solar
metallicity model, both at 8 Gyr.
\end{table*}

\begin{table*}[p]\footnotesize
\addtocounter{table}{-1}
\caption{\label{tab:slickellip2}}
\begin{center}
{\sc sLick indices on Average Spectra\\}
%%% \begin{tabular}{cc cccccccc} 
\begin{tabular}{ccccccccc}
\tableskip\tableline\tableline\tableskip
Magnitude & Environment/  & Fe4383 & Ca4455 & Fe4531 & Fe4668 & H$\beta$ & Mg${}_2$ & Mg~$b$\\
& Redshift & \AA & \AA & \AA & \AA & \AA & mag & \AA\\
\tableskip\tableline\tableline\tableskip
$-22.5<M_r<-22.0$ & All
 & $4.158$ (25) & $1.017$ (12) & $2.975$ (19) & $6.377$ (31) & $1.583$ (13) & $0.2570$ (2) & $3.781$ (14)\\
$-22.0<M_r<-21.5$ & All
 & $4.039$ (18) & $1.005$ (8) & $2.959$ (11) & $6.121$ (17) & $1.575$ (9) & $0.2441$ (2) & $3.684$ (12)\\
$-21.5<M_r<-21.0$ & All
 & $3.959$ (18) & $0.976$ (8) & $2.918$ (11) & $5.774$ (20) & $1.497$ (8) & $0.2345$ (2) & $3.543$ (12)\\
$-21.0<M_r<-20.5$ & All
 & $3.798$ (25) & $0.933$ (7) & $2.856$ (14) & $5.372$ (17) & $1.472$ (7) & $0.2265$ (2) & $3.382$ (13)\\
\tableskip\tableline\tableskip
Slope per mag & 
 & {\bf +0.226} & {\bf +0.061} & {\bf +0.085} & {\bf +0.704} & {\bf +0.086} & {\bf +0.0202} & {\bf +0.270}\\
Error & 
 & $0.022$ & $0.008$ & $0.014$ & $0.020$ & $0.009$ & $0.0003$ & $0.012$\\
\tableskip\tableline\tableline\tableskip
$-22.5<M_r<-22.0$ & $N\ge10$
 & $4.251$ (46) & $1.099$ (22) & $3.002$ (40) & $6.450$ (63) & $1.512$ (26) & $0.2685$ (7) & $4.006$ (28)\\
$-22.5<M_r<-22.0$ & $4\le N\le9$
 & $4.133$ (40) & $1.020$ (18) & $2.985$ (31) & $6.324$ (41) & $1.578$ (21) & $0.2552$ (5) & $3.770$ (22)\\
$-22.5<M_r<-22.0$ & $0\le N\le3$
 & $4.062$ (70) & $0.977$ (32) & $2.980$ (53) & $6.311$ (99) & $1.674$ (38) & $0.2477$ (10) & $3.678$ (42)\\
\tableskip\tableline\tableskip
$-22.5<M_r<-22.0$ & $N$ slope
 & +0.198 & {\bf +0.128} & +0.024 & +0.171 & {\boldmath $-0.156$} & {\bf +0.0220} & {\bf +0.358}\\
 & Error
 & $0.081$ & $0.038$ & $0.066$ & $0.111$ & $0.045$ & $0.0012$ & $0.049$\\
\tableskip\tableline\tableline\tableskip
$-22.0<M_r<-21.5$ & $N\ge10$
 & $4.119$ (35) & $1.030$ (13) & $3.010$ (24) & $6.284$ (46) & $1.521$ (25) & $0.2559$ (7) & $3.801$ (31)\\
$-22.0<M_r<-21.5$ & $4\le N\le9$
 & $3.988$ (29) & $0.981$ (11) & $2.932$ (20) & $6.072$ (34) & $1.537$ (13) & $0.2441$ (2) & $3.655$ (12)\\
$-22.0<M_r<-21.5$ & $0\le N\le3$
 & $3.976$ (44) & $0.992$ (16) & $2.898$ (30) & $5.952$ (39) & $1.601$ (23) & $0.2351$ (5) & $3.529$ (21)\\
\tableskip\tableline\tableskip
$-22.0<M_r<-21.5$ & $N$ slope
 & +0.158 & +0.043 & +0.116 & {\bf +0.325} & $-0.083$ & {\bf +0.0202} & {\bf +0.268}\\
 & Error
 & $0.056$ & $0.021$ & $0.039$ & $0.061$ & $0.034$ & $0.0008$ & $0.036$\\
\tableskip\tableline\tableline\tableskip
$-21.5<M_r<-21.0$ & $N\ge10$
 & $4.006$ (41) & $0.986$ (18) & $2.932$ (26) & $5.881$ (43) & $1.542$ (25) & $0.2411$ (8) & $3.619$ (32)\\
$-21.5<M_r<-21.0$ & $4\le N\le9$
 & $3.937$ (37) & $0.959$ (14) & $2.891$ (20) & $5.761$ (34) & $1.529$ (16) & $0.2368$ (5) & $3.556$ (18)\\
$-21.5<M_r<-21.0$ & $0\le N\le3$
 & $3.872$ (30) & $0.956$ (15) & $2.874$ (20) & $5.682$ (36) & $1.493$ (12) & $0.2313$ (4) & $3.437$ (15)\\
\tableskip\tableline\tableskip
$-21.5<M_r<-21.0$ & $N$ slope
 & +0.134 & +0.029 & +0.055 & {\bf +0.196} & +0.054 & {\bf +0.0101} & {\bf +0.199}\\
 & Error
 & $0.051$ & $0.024$ & $0.033$ & $0.057$ & $0.026$ & $0.0008$ & $0.032$\\
\tableskip\tableline\tableline\tableskip
$-21.0<M_r<-20.5$ & $N\ge10$
 & $3.889$ (60) & $0.957$ (28) & $2.903$ (42) & $5.535$ (61) & $1.519$ (27) & $0.2348$ (11) & $3.490$ (51)\\
$-21.0<M_r<-20.5$ & $4\le N\le9$
 & $3.827$ (34) & $0.966$ (10) & $2.879$ (25) & $5.322$ (36) & $1.452$ (16) & $0.2282$ (5) & $3.420$ (22)\\
$-21.0<M_r<-20.5$ & $0\le N\le3$
 & $3.719$ (41) & $0.907$ (16) & $2.834$ (23) & $5.335$ (43) & $1.474$ (19) & $0.2233$ (5) & $3.313$ (22)\\
\tableskip\tableline\tableskip
$-21.0<M_r<-20.5$ & $N$ slope
 & +0.179 & +0.073 & +0.074 & +0.160 & +0.030 & {\bf +0.0107} & {\bf +0.192}\\
 & Error
 & $0.071$ & $0.029$ & $0.045$ & $0.073$ & $0.033$ & $0.0011$ & $0.048$\\
\tableskip\tableline\tableline\tableskip
$M_g<-21.8$ & $0.15<z<0.25$
 & $4.149$ (70) & $1.041$ (29) & $3.016$ (49) & $6.663$ (82) & $1.607$ (40) & $0.2670$ (8) & $3.934$ (34)\\
$M_g<-21.8$ & $0.25<z<0.35$
 & $4.157$ (78) & $1.050$ (38) & $2.952$ (51) & $6.334$ (81) & $1.595$ (28) & $0.2622$ (8) & $3.863$ (35)\\
$M_g<-21.8$ & $0.35<z<0.45$
 & $4.004$ (73) & $0.991$ (32) & $2.891$ (42) & $6.045$ (66) & $1.713$ (22) & $0.2423$ (11) & $3.608$ (43)\\
$M_g<-21.8$ & $0.45<z<0.50$
 & $3.634$ (91) & $0.911$ (45) & $2.802$ (77) & $5.954$ (109) & $1.810$ (46) & $0.2287$ (21) & $3.418$ (86)\\
\tableskip\tableline\tableskip
 & Slope per $\Delta z =0.1$
 & {\boldmath $-0.152$} & $-0.039$ & $-0.068$ & {\boldmath $-0.259$} & {\bf +0.076} & {\boldmath $-0.0123$} & {\boldmath $-0.166$}\\
 & Error
 & $0.036$ & $0.016$ & $0.026$ & $0.040$ & $0.018$ & $0.0006$ & $0.023$\\
\tableskip\tableline\tableline\tableskip
Model: 8 Gyr & [Z/H]=0.0
 & $4.373$ & \nodata  & \nodata  & \nodata  & $1.804$ & $0.2189$ & $3.134$\\
12 Gyr - 8 Gyr & [Z/H]=0.0
 & $0.672$ & \nodata  & \nodata  & \nodata  & $-0.212$ & $0.0314$ & $0.364$\\
8 Gyr & [Z/H]=0.2 - solar
 & $0.777$ & \nodata  & \nodata  & \nodata  & $-0.007$ & $0.0359$ & $0.381$\\
\tableskip\tableline\tableline\tableskip
\end{tabular}

%%% \end{tabular}
\end{center}
NOTES.---%
Continuation of Table \protect\ref{tab:slickellip1}.
\end{table*}

\begin{table*}[p]\footnotesize
\addtocounter{table}{-1}
\caption{\label{tab:slickellip3}}
\begin{center}
{\sc sLick indices on Average Spectra\\}
%%% \begin{tabular}{cc cccccccc} 
\begin{tabular}{ccccccccc}
\tableskip\tableline\tableline\tableskip
Magnitude & Environment/  & Fe5270 & Fe5335 & Fe5406 & Fe5709 & Fe5782 & Na~D & TiO${}_2$\\
& Redshift & \AA & \AA & \AA & \AA & \AA & \AA & mag\\
\tableskip\tableline\tableline\tableskip
$-22.5<M_r<-22.0$ & All
 & $2.340$ (18) & $1.873$ (19) & $1.246$ (11) & $0.744$ (7) & $0.594$ (8) & $3.687$ (13) & $0.0745$ (2)\\
$-22.0<M_r<-21.5$ & All
 & $2.380$ (11) & $1.945$ (11) & $1.273$ (6) & $0.765$ (5) & $0.597$ (4) & $3.493$ (7) & $0.0723$ (2)\\
$-21.5<M_r<-21.0$ & All
 & $2.348$ (10) & $1.919$ (8) & $1.263$ (5) & $0.771$ (5) & $0.587$ (3) & $3.295$ (5) & $0.0709$ (2)\\
$-21.0<M_r<-20.5$ & All
 & $2.292$ (13) & $1.856$ (14) & $1.228$ (9) & $0.769$ (6) & $0.573$ (4) & $3.096$ (7) & $0.0701$ (1)\\
\tableskip\tableline\tableskip
Slope per mag & 
 & {\bf +0.050} & +0.031 & +0.017 & $-0.014$ & {\bf +0.018} & {\bf +0.395} & {\bf +0.0027}\\
Error & 
 & $0.013$ & $0.014$ & $0.008$ & $0.006$ & $0.005$ & $0.008$ & $0.0001$\\
\tableskip\tableline\tableline\tableskip
$-22.5<M_r<-22.0$ & $N\ge10$
 & $2.361$ (25) & $1.924$ (26) & $1.274$ (18) & $0.755$ (15) & $0.625$ (14) & $3.934$ (32) & $0.0778$ (4)\\
$-22.5<M_r<-22.0$ & $4\le N\le9$
 & $2.341$ (25) & $1.911$ (27) & $1.248$ (20) & $0.757$ (9) & $0.600$ (13) & $3.708$ (18) & $0.0744$ (4)\\
$-22.5<M_r<-22.0$ & $0\le N\le3$
 & $2.374$ (37) & $1.889$ (39) & $1.261$ (34) & $0.733$ (15) & $0.594$ (16) & $3.583$ (24) & $0.0729$ (8)\\
\tableskip\tableline\tableskip
$-22.5<M_r<-22.0$ & $N$ slope
 & $-0.005$ & +0.034 & +0.023 & +0.022 & +0.033 & {\bf +0.336} & {\bf +0.0055}\\
 & Error
 & $0.043$ & $0.046$ & $0.036$ & $0.022$ & $0.022$ & $0.040$ & $0.0008$\\
\tableskip\tableline\tableline\tableskip
$-22.0<M_r<-21.5$ & $N\ge10$
 & $2.398$ (27) & $1.950$ (28) & $1.284$ (24) & $0.746$ (10) & $0.588$ (11) & $3.678$ (21) & $0.0759$ (4)\\
$-22.0<M_r<-21.5$ & $4\le N\le9$
 & $2.334$ (14) & $1.913$ (15) & $1.244$ (12) & $0.763$ (7) & $0.593$ (8) & $3.449$ (11) & $0.0723$ (2)\\
$-22.0<M_r<-21.5$ & $0\le N\le3$
 & $2.323$ (19) & $1.930$ (26) & $1.223$ (16) & $0.761$ (10) & $0.583$ (10) & $3.363$ (17) & $0.0699$ (2)\\
\tableskip\tableline\tableskip
$-22.0<M_r<-21.5$ & $N$ slope
 & +0.064 & +0.017 & +0.057 & $-0.015$ & +0.006 & {\bf +0.300} & {\bf +0.0056}\\
 & Error
 & $0.032$ & $0.038$ & $0.028$ & $0.014$ & $0.015$ & $0.027$ & $0.0004$\\
\tableskip\tableline\tableline\tableskip
$-21.5<M_r<-21.0$ & $N\ge10$
 & $2.390$ (36) & $1.914$ (31) & $1.269$ (21) & $0.755$ (15) & $0.585$ (12) & $3.373$ (16) & $0.0728$ (4)\\
$-21.5<M_r<-21.0$ & $4\le N\le9$
 & $2.315$ (15) & $1.875$ (20) & $1.235$ (15) & $0.760$ (9) & $0.577$ (6) & $3.283$ (10) & $0.0709$ (2)\\
$-21.5<M_r<-21.0$ & $0\le N\le3$
 & $2.305$ (16) & $1.869$ (18) & $1.217$ (13) & $0.755$ (11) & $0.567$ (6) & $3.251$ (12) & $0.0701$ (2)\\
\tableskip\tableline\tableskip
$-21.5<M_r<-21.0$ & $N$ slope
 & +0.057 & +0.038 & +0.048 & +0.001 & +0.018 & {\bf +0.114} & {\bf +0.0026}\\
 & Error
 & $0.034$ & $0.035$ & $0.025$ & $0.019$ & $0.013$ & $0.021$ & $0.0005$\\
\tableskip\tableline\tableline\tableskip
$-21.0<M_r<-20.5$ & $N\ge10$
 & $2.318$ (46) & $1.846$ (30) & $1.235$ (24) & $0.751$ (20) & $0.566$ (15) & $3.170$ (25) & $0.0729$ (4)\\
$-21.0<M_r<-20.5$ & $4\le N\le9$
 & $2.315$ (21) & $1.859$ (22) & $1.243$ (16) & $0.773$ (11) & $0.569$ (9) & $3.122$ (10) & $0.0699$ (2)\\
$-21.0<M_r<-20.5$ & $0\le N\le3$
 & $2.226$ (28) & $1.838$ (30) & $1.225$ (17) & $0.758$ (14) & $0.562$ (9) & $3.046$ (14) & $0.0694$ (4)\\
\tableskip\tableline\tableskip
$-21.0<M_r<-20.5$ & $N$ slope
 & +0.117 & +0.007 & +0.014 & $-0.001$ & +0.006 & {\bf +0.133} & {\bf +0.0035}\\
 & Error
 & $0.051$ & $0.044$ & $0.029$ & $0.025$ & $0.017$ & $0.027$ & $0.0006$\\
\tableskip\tableline\tableline\tableskip
$M_g<-21.8$ & $0.15<z<0.25$
 & $2.386$ (31) & $1.967$ (36) & $1.265$ (26) & $0.739$ (24) & $0.576$ (22) & $3.952$ (33) & \nodata \\
$M_g<-21.8$ & $0.25<z<0.35$
 & $2.353$ (30) & $1.890$ (33) & $1.254$ (29) & $0.709$ (22) & $0.628$ (24) & $3.872$ (40) & \nodata \\
$M_g<-21.8$ & $0.35<z<0.45$
 & $2.325$ (46) & $1.834$ (49) & $1.191$ (32) & $0.627$ (25) & $0.555$ (26) & $3.614$ (60) & \nodata \\
$M_g<-21.8$ & $0.45<z<0.50$
 & $2.127$ (105) & $1.762$ (107) & $1.153$ (59) & $0.764$ (94) & $0.579$ (77) & $3.405$ (142) & \nodata \\
\tableskip\tableline\tableskip
 & Slope per $\Delta z =0.1$
 & $-0.048$ & $-0.067$ & $-0.037$ & $-0.044$ & $-0.007$ & {\boldmath $-0.160$} & \nodata \\
 & Error
 & $0.024$ & $0.026$ & $0.017$ & $0.017$ & $0.016$ & $0.029$ & \nodata \\
\tableskip\tableline\tableline\tableskip
Model: 8 Gyr & [Z/H]=0.0
 & $2.598$ & $1.907$ & $1.295$ & \nodata  & \nodata  & \nodata  & \nodata \\
12 Gyr - 8 Gyr & [Z/H]=0.0
 & $0.259$ & $0.220$ & $0.157$ & \nodata  & \nodata  & \nodata  & \nodata \\
8 Gyr & [Z/H]=0.2 - solar
 & $0.335$ & $0.293$ & $0.217$ & \nodata  & \nodata  & \nodata  & \nodata \\
\tableskip\tableline\tableline\tableskip
\end{tabular}

%%% \end{tabular}
\end{center}
NOTES.---%
Continuation of Table \protect\ref{tab:slickellip1}.
\end{table*}

\end{document}